\documentclass[reprint,amsmath,amssymb,aps,nofootinbib,twocolumn]{revtex4-2}
\pdfoutput=1
\usepackage{graphicx,amsmath,amssymb,amsthm}
\usepackage[hidelinks]{hyperref}
\usepackage{xcolor}
\usepackage{ulem}
\usepackage{thmtools, thm-restate}
\usepackage{epsfig}
\usepackage{epstopdf}
\usepackage{latexsym}
\usepackage{graphicx}
\usepackage{booktabs}
\usepackage{bbm}
\usepackage{color}
\usepackage{physics}
\usepackage{tensor}
\usepackage{verbatim}
\usepackage[caption=false]{subfig}
\usepackage{tikz}
\usepackage{bigints}
\usepackage{ifthen}
\usetikzlibrary{matrix}
\usetikzlibrary{decorations.markings,calc,shapes,decorations.pathmorphing}
\usetikzlibrary{patterns}
\usetikzlibrary{positioning}
\usepackage{textpos}

\usepackage{hyperref}

\usepackage{xcolor}

\hypersetup{
colorlinks,
linkcolor={blue!80!black},
citecolor={blue!80!black},
urlcolor={blue!80!black},
linktoc=page
}

\newcommand\beq{\begin{equation}}
\newcommand\eeq{\end{equation}}

\newcommand{\be}{\begin{equation}}
\newcommand{\ee}{\end{equation}}

\begin{document}

\title{Trans-IR Flows to Black Hole Singularities}


\author{Elena Caceres$^{a}$}
\email{elenac@utexas.edu}
\author{Arnab Kundu$^{b}$}
\email{arnab.kundu@gmail.com}
\author{Ayan K.~Patra$^{b}$}
\email{ayankumarpatra@gmail.com}
\author{Sanjit Shashi$^{a}$}
\email{sshashi@utexas.edu}
\affiliation{$^a$Theory Group, Department of Physics, University of Texas, Austin, Texas 78712, USA.}
\affiliation{$^b$Theory Division, Saha Institute of Nuclear Physics, HBNI, 1/AF Bidhannagar, Kolkata 700064, India.}

\preprint{\today}
\begin{abstract}
\noindent We study analytic continuations of holographic renormalization group (RG) flows beyond their infrared (IR) fixed points. Such ``trans-IR" flows are a natural framework for describing physics inside of black holes. First, we construct a monotonic holographic $a$-function which counts degrees of freedom along a trans-IR flow. Using this function, we argue that the degrees of freedom ``thin out" and vanish when flowing to a trans-IR endpoint, represented by a Kasner singularity. We then recast well-studied quantum information probes in the language of trans-IR flows, finding that entanglement and complexity from volume generally fail to fully probe the trans-IR while 2-point correlations and complexity from action generally do so in a complementary manner.
\end{abstract}

\maketitle



\section{Introduction}\label{sec1}

A ``trans-IR" flow is constructed by analytically continuing a renormalization group (RG) flow beyond its IR fixed point to complex energy scales. In holographic setups where an RG flow is simply gravity, there is a natural interpretation of following the trans-IR flow of a UV thermal state---accessing a black hole's interior by flowing towards the singularity \cite{Frenkel:2020ysx}. That analytic continuation is needed to sensibly relate boundary field-theoretic data to bulk geometry probing the interior is well-known in the AdS/CFT correspondence \cite{Maldacena:1997re,Fidkowski:2003nf,Festuccia:2005pi,Balasubramanian:2019qwk,Grinberg:2020fdj}. In this spirit, we propose that the language of trans-IR flows is a natural framework for discussing physics inside of black holes.

From bulk metric functions, we first construct a ``thermal" analog to more conventional holographic $a$-functions \cite{Henningson:1998gx,Freedman:1999gp}, calling it $a_T$. While $a_T$ is stationary at both the boundary (the UV) and the horizon (the IR), it is still monotonic even along the trans-IR flow, thus satisfying an extension of the holographic $a$-theorem \cite{Myers:2010xs,Myers:2010tj} to imaginary energy scales. When looking to ``free Kasner flows" \cite{Frenkel:2020ysx,Caceres:2021fuw}, we find $a_T \to 0$ at the singularity (the trans-IR endpoint). In other words, all degrees of freedom are lost at the singularity. We then discuss quantum information in the context of trans-IR flows corresponding to static black holes. Entanglement \cite{Ryu:2006bv,Hubeny:2007xt} and complexity from volume \cite{Susskind:2014rva,Stanford:2014jda,Susskind:2014moa} only probe the trans-IR partially, but 2-point correlations \cite{Balasubramanian:1999zv,Fidkowski:2003nf} and complexity from action \cite{Brown:2015bva,Brown:2015lvg} probe it fully.

Let us briefly review \textit{holographic RG flow} \cite{Balasubramanian:1999jd}. The bulk ``radial" extra dimension $\rho$ of AdS is treated as an energy scale parameterizing an RG flow from a UV conformal field theory (CFT) on the boundary to an IR field theory deep in the bulk. The bulk gravitational dynamics are also the dynamics of this flow \cite{deBoer:1999tgo,deBoer:2000cz,Bianchi:2001kw,Fukuma:2002sb,Papadimitriou:2004ap,Papadimitriou:2005ii}. We trigger RG flows by adding dynamical bulk fields which are dual to relevant operators on the boundary \cite{Aharony:1999ti}. For example, take the deformation
\begin{equation}
I_{\mathcal{O}} = \int d^dx\, \phi_0 \mathcal{O}.\label{scalarDeformAct}
\end{equation}
$\mathcal{O}$ is a relevant scalar operator and $\phi_0$ is a source. In the bulk, we have a gravity + scalar theory. For Einstein gravity, these flows have been studied extensively \cite{Bourdier:2013axa,Kiritsis:2014kua,Kiritsis:2016kog,Gursoy:2018umf}.

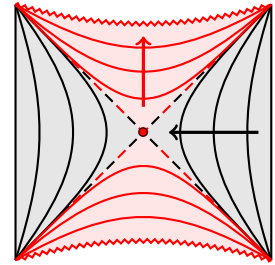
\begin{figure}
\centering
\begin{tikzpicture}[scale=1.7]
\draw[-,draw=none,fill=black!10] (1,1) to (1,-1) to (0,0) to (1,1); 
\draw[-,draw=none,fill=black!10] (-1,1) to (-1,-1) to (0,0) to (-1,1);

\draw[-,draw=none,fill=red!10] (-1,1) ..controls (-0.3,0.8) and (0.3,0.8) .. (1,1) to (0,0) to (-1,1); 
\draw[-,draw=none,fill=red!10] (1,-1) ..controls (0.3,-0.8) and (-0.3,-0.8) .. (-1,-1) to (0,0) to (1,-1);

\draw[-,thick,red,decoration = {zigzag,segment length = 1mm, amplitude = 0.25mm},decorate] (-1,1) ..controls (-0.3,0.8) and (0.3,0.8) .. (1,1);
\draw[-,thick] (1,1) to (1,-1);
\draw[-,thick,red,decoration = {zigzag,segment length = 1mm, amplitude = 0.25mm},decorate] (1,-1) ..controls (0.3,-0.8) and (-0.3,-0.8) .. (-1,-1);
\draw[-,thick] (-1,-1) to (-1,1);

\draw[-,dashed,thick,dash pattern= on 4pt off 8pt,dash phase=6pt,red] (-1,1) to (1,-1);
\draw[-,dashed,thick,dash pattern= on 4pt off 8pt,dash phase=6pt,red] (1,1) to (-1,-1);

\draw[-,dashed,thick,dash pattern= on 4pt off 8pt] (-1,1) to (1,-1);
\draw[-,dashed,thick,dash pattern= on 4pt off 8pt] (1,1) to (-1,-1);

\draw[-,thick] (1,1) .. controls (0.75,0.2) and (0.75,-0.2) .. (1,-1);
\draw[-,thick] (1,1) .. controls (0.4,0.2) and (0.4,-0.2) .. (1,-1);
\draw[-,thick] (1,1) .. controls (0.05,0.05) and (0.05,-0.05) .. (1,-1);

\draw[-,thick] (-1,1) .. controls (-0.75,0.2) and (-0.75,-0.2) .. (-1,-1);
\draw[-,thick] (-1,1) .. controls (-0.4,0.2) and (-0.4,-0.2) .. (-1,-1);
\draw[-,thick] (-1,1) .. controls (-0.05,0.05) and (-0.05,-0.05) .. (-1,-1);

\draw[-,thick,red] (-1,1) .. controls (-0.2,0.55) and (0.2,0.55) .. (1,1);
\draw[-,thick,red] (-1,1) .. controls (-0.2,0.3) and (0.2,0.3) .. (1,1);
\draw[-,thick,red] (-1,1) .. controls (-0.02,0.0175) and (0.02,0.0175) .. (1,1);

\draw[-,thick,red] (-1,-1) .. controls (-0.2,-0.55) and (0.2,-0.55) .. (1,-1);
\draw[-,thick,red] (-1,-1) .. controls (-0.2,-0.3) and (0.2,-0.3) .. (1,-1);
\draw[-,thick,red] (-1,-1) .. controls (-0.02,-0.0175) and (0.02,-0.0175) .. (1,-1);

\node[red] at (0,0) {$\bullet$};
\node at (0,0) {$\circ$};

\draw[->,very thick] (0.9,0) to (0.2,0);

\draw[->,very thick,red] (0,0.2) to (0,0.75);
\end{tikzpicture}
\caption{A two-sided asymptotically anti-de Sitter (AdS) black hole, with the exterior in gray and the interior in red. The lines are constant radial slices, with the horizon being the dashed lines. The black arrow represents a conventional UV $\to$ IR holographic RG flow while the red arrow indicates a trans-IR flow parameterized by a timelike radial coordinate.}
\label{figs:holoRGFlow}
\end{figure}

This approach naturally yields field-theoretic trans-IR states. There are gravitational solutions for which $\rho$ changes signature from spacelike to timelike---black holes (Figure \ref{figs:holoRGFlow}). Thus, timelike ``scaling" from the horizon to the singularity is recognized as a trans-IR flow. Such flows have been studied for scalar deformations \cite{Frenkel:2020ysx}; the black hole interiors behave as Kasner universes \cite{Kasner:1921zz,Belinskii:1973zz,Das:2006dz}.

So, we may understand black hole interiors as trans-IR flows. However, what does it mean to follow a trans-IR flow in the language of field theory? We address this broad question by focusing on three specific questions:
\begin{itemize}
\item[\textbf{(Q1)}] Do trans-IR flows obey a monotonicity condition? (Section \ref{sec:Q1})

\item[\textbf{(Q2)}] What happens to the degrees of freedom near the endpoint of a trans-IR flow? (Section \ref{sec:Q2})

\item[\textbf{(Q3)}] How does quantum information of the UV state encode the trans-IR flow? (Section \ref{sec:Q3})
\end{itemize}
Using holography, we answer each of these in order. Ancillary discussion regarding time in the interior (Appendix \ref{app:A}), extremal surfaces (Appendix \ref{app:B}), and quantum information probes of the trans-IR regime (Appendices \ref{app:C} and \ref{app:D}) is left to the appendices.

\section{The Monotonic ${\boldsymbol{a_T}}$-Function}\label{sec:Q1}

\subsection{Defining the Function}

We first construct the thermal generalization of the $a$-function of \cite{Freedman:1999gp,Myers:2010xs,Myers:2010tj} in Einstein gravity with negative cosmological constant. Take a domain-wall ansatz,
\begin{equation}
ds^2 = e^{2A(\rho)}\left[-f(\rho)^2 dt^2 + d\vec{x}^2\right] + d\rho^2,
\end{equation}
with $t \in \mathbb{R}$, $\vec{x} \in \mathbb{R}^{d-1}$, $\rho \geq 0$. $A(\rho)$ and $f(\rho)$ are generally arbitrary functions. $\rho$ is the energy scale, with the conformal boundary $\rho \to \infty$ representing the UV.

Setting $f(\rho) = 1$ yields a domain wall with flat slicing \cite{Freedman:1999gp}. We then get vacuum AdS$_{d+1}$ with radius $\ell$ when $A(\rho) = \rho/\ell$.\footnote{In most of the paper, we will just set $\ell = 1$. However, we retain factors of $\ell$ in this subsection as they are necessary to construct the function of interest.} It is immediate that $A'(\rho) = 1/\ell$, so we may introduce $\rho$ dependence to the $a$ central charge.\footnote{Early AdS/CFT literature \cite{Henningson:1998gx,Freedman:1999gp} identified this as the coefficient of the even-$d$ trace anomaly, but \cite{Myers:2010xs,Myers:2010tj} later noted that $a$ appears for any $d$ in entanglement entropy.}
\begin{align}
&a = \frac{\pi^{d/2}}{\Gamma\left(\frac{d}{2}\right)} \left(\frac{\ell}{\ell_P}\right)^{d-1}\\
&\implies a(\rho) = \frac{\pi^{d/2}}{\Gamma\left(\frac{d}{2}\right)\ell_P^{d-1}} \left[\frac{1}{A'(\rho)}\right]^{d-1}.\label{aFunction1}
\end{align}
For nontrivial holographic RG flows with $A(\rho) \sim \rho/\ell$ only near the fixed points, not only does \eqref{aFunction1} become nonconstant but it can also become monotonic ($da/d\rho \geq 0$). Monotonicity is achieved if the matter sourcing the RG flow satisfies the null energy condition (NEC) \cite{Freedman:1999gp,Myers:2010xs,Myers:2010tj}---for any null vector $k^\alpha$ and the bulk stress tensor $T_{\mu\nu}$,
\begin{equation}
k^\mu k^\nu T_{\mu\nu} \geq 0.\label{NEC}
\end{equation}
However, we are concerned with black holes, so we take $f(\rho) > 0$ with a horizon at $\rho = 0$, i.e. $f(\rho) = f_1\rho + O(\rho^3)$. We get the AdS-Schwarzschild solution\footnote{Note that the temperature of this black hole is set to $1/(2\pi \ell)$.} when \cite{Hartman:2013qma}
\begin{equation}
e^{A(\rho)} = \frac{2}{d}\cosh\left(\frac{d \rho}{2\ell}\right)^{2/d},\ \ f(\rho) = \tanh\left(\frac{d \rho}{2\ell}\right).\label{adsSchwarz}
\end{equation}
This time, we use that $\ell = f(\rho)/A'(\rho)$ to write the extension of \eqref{aFunction1} allowing finite temperatures as
\begin{equation}
a_T(\rho) = \frac{\pi^{d/2}}{\Gamma\left(\frac{d}{2}\right)\ell_P^{d-1}}\left[\frac{f(\rho)}{A'(\rho)}\right]^{d-1}.\label{aFunction2}
\end{equation}
$a_T$ for AdS-Schwarzschild matches $a$ for empty AdS$_{d+1}$---a consequence of both geometries being dual to states of the same UV CFT. This is true even in the interior accessed by analytic continuation of \eqref{adsSchwarz} to
\begin{equation}
\rho = i\kappa,\ \ t = t_I - \text{sgn}(t_I)\frac{i\gamma}{2T},\label{analyticCont}
\end{equation}
where $\kappa > 0$, $t_I \in \mathbb{R}$, $\gamma$ is a half integer,\footnote{We elaborate on $\gamma$ in Appendix \ref{app:A}, but for now it suffices to note that it depends on how we choose to analytically continue coordinate time into the interior, and that past work \cite{Fidkowski:2003nf,Hartman:2013qma,Frenkel:2020ysx} typically has taken $|\gamma| = 1/2$.} and
\begin{equation}
T = \frac{e^{A(0)} f_1}{2\pi}
\end{equation}
is the black hole temperature.

\eqref{analyticCont} describes the trans-IR, so we answer (Q1) by exploring $a_T$ in this coordinate domain.

\subsection{Proof of Monotonicity}

The usual procedure to confirm monotonicity of $a_T$ is to analyze its derivative along the flow,
\begin{equation}
\begin{split}
\frac{da_T}{d\rho}
&= \frac{(d-1)\pi^{d/2}}{\Gamma\left(\frac{d}{2}\right)\ell_P^{d-1}}\frac{f(\rho)^{d-2}}{A'(\rho)^d}\\
&\qquad\times \left[f'(\rho)A'(\rho)-f(\rho)A''(\rho)\right].
\end{split}
\end{equation}
The general Einstein equations $G_{\mu\nu} + \Lambda g_{\mu\nu} = \ell_P^{d-1} T_{\mu\nu}$ (now setting $\ell = 1$, so $\Lambda = -d(d-1)/2$) imply that
\begin{align}
&\tensor{T}{^\rho_\rho} - \tensor{T}{^t_t} = \frac{(d-1)}{\ell_P^{d-1} f(\rho)} \left[f'(\rho) A'(\rho) - f(\rho)A''(\rho)\right]\\
&\implies \frac{da_T}{d\rho} = \frac{\pi^{d/2}}{\Gamma\left(\frac{d}{2}\right) f(\rho)} \left[\frac{f(\rho)}{A'(\rho)}\right]^d \left(\tensor{T}{^\rho_\rho}-\tensor{T}{^t_t}\right).
\end{align}
Using the null vector $\vec{k} = [e^{-A(\rho)}/f(\rho)]\partial_t + \partial_\rho$, we can write the NEC \eqref{NEC} as
\begin{equation}
\tensor{T}{^\rho_\rho} - \tensor{T}{^t_t} \geq 0.\label{NECspec}
\end{equation}
\eqref{NECspec} holds everywhere and, in conjunction with the positivity of $A'(\rho)$ near the boundary, implies that $a_T$ decreases along the flow towards the horizon.\footnote{This reasoning is analogous to the monotonicity argument for the $a$-function that \cite{Myers:2010xs,Myers:2010tj} construct in Gauss-Bonnet gravity.} However, proving monotonicity everywhere requires separate treatment of the interior. In fact, even checking that the horizon is the IR fixed point is problematic in these coordinates because finiteness of $da_T/d\rho$ is not obvious.

For these reasons, we find it easier to prove monotonicity in another set of coordinates where the exterior and interior are connected along a real radial coordinate. We transform $\rho \to r$ (with $r \in (0,\infty)$) by setting
\begin{equation}
e^{2A(\rho)} = \frac{1}{r^2},\ \ f(\rho)^2 = F(r) e^{-\chi(r)},\ \ \frac{dr}{d\rho} = -r\sqrt{F(r)}.\label{coordTrans}
\end{equation}
Here, $\chi(r)$ is an analytic function of $r$ with $\chi(0) = 0$, while $F(r)$ must have a simple root $r = r_h$---the horizon radius in the $r$ slicing.\footnote{This reparameterization rescales the black hole temperature. It now depends on $r_h$.} Furthermore $r < r_h$ is the exterior ($F(r) > 0$), while $r > r_h$ is the interior ($F(r) < 0$). The resulting metric is the Schwarzschild-like one of \cite{Frenkel:2020ysx,Caceres:2021fuw}
\begin{equation}
ds^2 = \frac{1}{r^2}\left[-F(r) e^{-\chi(r)} dt^2 + \frac{dr^2}{F(r)} + d\vec{x}^2\right],\label{schwarzLike}
\end{equation}
which has temperature,
\begin{equation}
T = \frac{|F'(r_h)| e^{-\chi(r_h)/2}}{4\pi}.\label{temp}
\end{equation}
We may also use \eqref{coordTrans} to rewrite \eqref{aFunction2} as
\begin{equation}
a_T(r) = \frac{\pi^{d/2}}{\Gamma\left(\frac{d}{2}\right)\ell_P^{d-1}} e^{-(d-1)\chi(r)/2}.\label{aFunctionSch}
\end{equation}
Then by differentiating and applying the Einstein equations in the Schwarzschild-like coordinates, we get
\begin{equation}
\frac{da_T}{dr} = -\frac{\pi^{d/2}}{\Gamma\left(\frac{d}{2}\right)} \frac{e^{-(d-1)\chi(r)/2}}{rF(r)^2} \left[F(r)\left(\tensor{T}{^r_r} - \tensor{T}{^t_t}\right)\right].\label{aDerivr}
\end{equation}
Now, we use $\vec{k} = e^{\chi(r)/2} \partial_t + F(r)\partial_r$ to write the NEC as
\begin{equation}
F(r)\left(\tensor{T}{^r_r} - \tensor{T}{^t_t}\right) \geq 0,
\end{equation}
which implies that $da_T/dr \leq 0$ everywhere.

Note that we are not done proving monotonicity. Strictly speaking, we care about whether $a_T$ monotonically decreases with respect to the energy scale $\rho$, even when we analytically continue to the trans-IR flow:
\begin{align}
\text{UV $\to$ IR $(r \leq r_h)$:}&\ \ \frac{da_T}{d\rho} \geq 0,\label{monoIneqExt}\\
\text{Trans-IR $(r \geq r_h)$:}&\ \ \frac{da_T}{d\kappa} \leq 0.\label{monoIneqInt}
\end{align}
\eqref{monoIneqExt} is immediate upon using the chain rule; in the exterior, $dr/d\rho \leq 0$. In fact, $dr/d\rho = 0$ at the horizon, and as $da_T/dr$ at the horizon is explicitly finite (being a combination of $\chi$ and $\chi'$), we have $da_T/d\rho|_{r=r_h} = 0$. This corroborates the horizon being the IR fixed point.

As for \eqref{monoIneqInt}, from \eqref{analyticCont} and \eqref{coordTrans},
\begin{equation}
r > r_h \implies \frac{1}{i}\frac{dr}{d\kappa} = -ir\sqrt{|F(r)|} \implies \frac{dr}{d\kappa} > 0.
\end{equation}
We thus conclude that $a_T$ indeed monotonically decreases both from the UV to the IR and along the trans-IR flow towards the singularity, answering (Q1).

\section{Losing Everything at the Singularity}\label{sec:Q2}

Holographically, the endpoint of a trans-IR flow from a UV thermal state is identified as the spacelike singularity of the corresponding black hole interior. Thus, we can answer (Q2) by analyzing how the near-singularity geometry is affected by backreaction. Describing backreaction is more easily done after specifying the matter sector, so we focus on the minimal case of \cite{Frenkel:2020ysx,Caceres:2021fuw}---Einstein gravity with a free massive scalar. In the resulting flows, we will find that $a_T \to 0$ at the singularity, around which the local geometry is that of a Kasner universe \cite{Kasner:1921zz,Belinskii:1973zz,Das:2006dz}. We thus call these geometries ``free Kasner flows."

It is reasonable to ask how to generalize from free Kasner flows. One way is to construct Kasner flows sourced by more complicated matter sectors \cite{Wang:2020nkd,Hartnoll:2020fhc,Sword:2021pfm,Mansoori:2021wxf,Das:2021vjf}, via numerical methods, and plot $a_T$---a concrete but \textit{ad hoc} approach. Another is to assume a Belinskii-Khalatnikov-Lifshitz (BKL) singularity \cite{Lifshitz:1963ps,Belinskii:1970ew,Belinskii:1982pk}---reviewed by \cite{Belinskii:2009wj}---as the trans-IR endpoint and analyze the near-singularity geometry. These approaches allow for exotic behavior, such as infinite sequences of distinct Kasner ``epochs" describing mixmaster dynamics \cite{Misner:1969hg,Damour:2002et,Damour:2002tc}.

\subsection{Free Kasner Flows}

The free-scalar-field matter sector of our theory is described by the action
\begin{equation}
I_S = -\frac{1}{4\ell_P^{d-1}}\int d^{d+1}x \sqrt{-g}\left(\nabla^\alpha \phi \nabla_\alpha \phi + m^2 \phi^2\right).
\end{equation}
The bulk equations of motion are the usual Klein-Gordon equation and the Einstein equations sourced by free scalar matter. For this theory, we will numerically determine the metric functions and compute $a_T$ explicitly. This is done in the Schwarzschild-like coordinates.

\subsubsection{Numerical Construction}

Consider a radial ansatz for the field $\phi = \phi(r)$ with $m^2 < 0$. This is dual to a constant relevant boundary scalar operator $\mathcal{O}$ which triggers an RG flow from the UV CFT through the deformation \eqref{scalarDeformAct}. Its conformal dimension $\Delta$ satisfies \cite{Aharony:1999ti}
\begin{equation}
m^2 = \Delta(\Delta - d).
\end{equation}
For this ansatz, the Klein-Gordon equation and the Einstein equations reduce to
\begin{align}
&\phi'' + \left(\frac{F'}{F} - \frac{d-1}{r} - \frac{\chi'}{2}\right)\phi' + \frac{\Delta(d-\Delta)}{r^2 F}\phi = 0,\label{eom1}\\
&\chi' - \frac{2F'}{F} - \frac{\Delta(d-\Delta)\phi^2}{(d-1)rF}-\frac{2d}{rF} + \frac{2d}{r} = 0,\label{eom2}\\
&\chi' - \frac{r}{d-1}(\phi')^2 = 0,\label{eom3}
\end{align}
where the primes denote derivatives with respect to $r$.

To solve \eqref{eom1}--\eqref{eom3} numerically, we expand $\{F,\chi,\phi\}$ around the horizon $r = r_h$, then perform a two-sided shooting method towards both the boundary $r = 0$ and the singularity $r = \infty$. The mathematical details are in \cite{Caceres:2021fuw}. The near-singularity geometry is a Kasner universe,
\begin{equation}
ds^2 \sim -d\tau^2 + \tau^{2p_t}dt^2 + \tau^{2p_x}d\vec{x}^2,\ \ \phi \sim -\sqrt{2}p_\phi \log\tau,
\end{equation}
where $\tau \in \mathbb{R}$ is a reparameterization of $r$, and $\{p_t,p_x,p_\phi\}$ are Kasner exponents satisfying
\begin{equation}
p_t + (d-1)p_x = 1,\ \ p_\phi^2 + p_t^2 + (d-1)p_x^2 = 1.
\end{equation}
We remark that some of the literature \cite{Damour:2002et,Damour:2002tc,Belinskii:2009wj} refers to this as ``Kasner-like" or ``generalized Kasner" geometry, because of the $p_\phi$. We will simply call it ``Kasner" for convenience.

\begin{figure}
\centering
\includegraphics[scale=0.85]{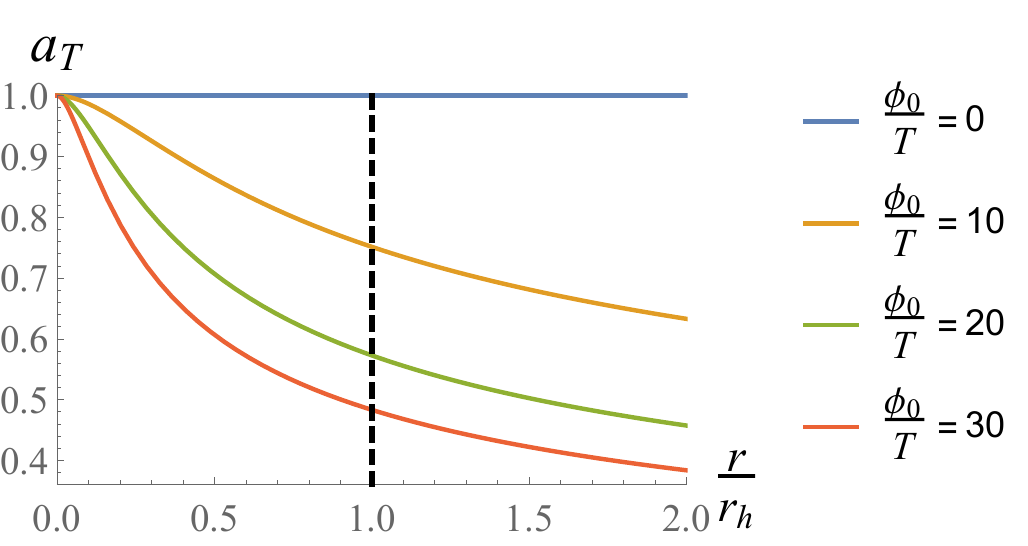}
\caption{The monotonic $a_T$-function (in units of $2\pi\ell_P^{-2}$) for a selection of free Kasner flows with $d = 3$, $\Delta = 2$ and as a function of the dimensionless ratio $r/r_h$. Each curve corresponds to a value of $\phi_0/T$. The dashed line is the horizon, with the trans-IR regime to its right.}
\label{figs:afunctionKasner}
\end{figure}

\begin{figure}
\centering
\includegraphics[scale=0.75]{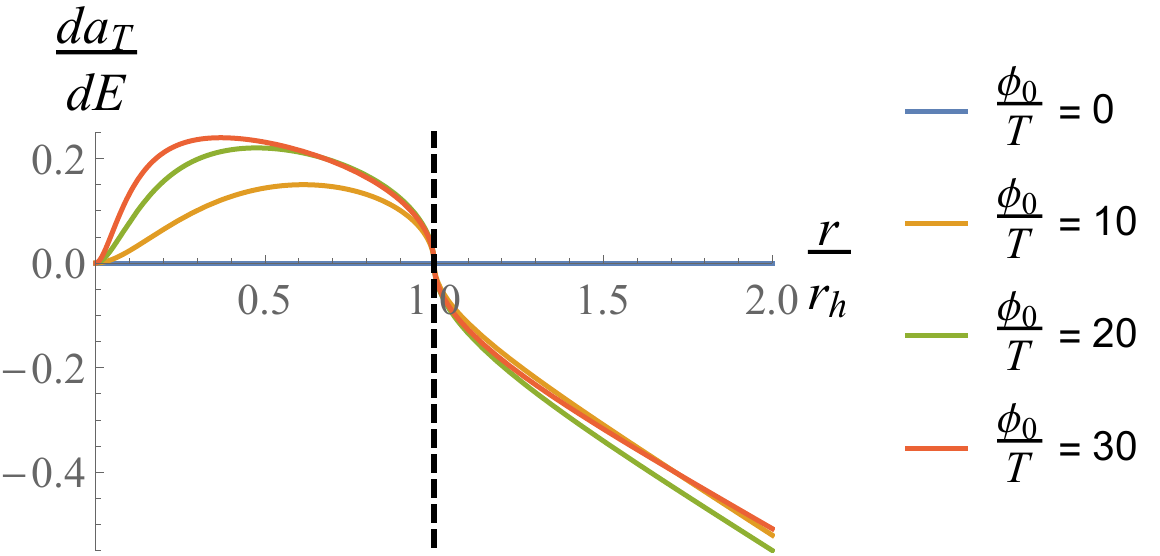}
\caption{The derivatives of $a_T$ (again in units of $2\pi\ell_P^{-2}$) along the free Kasner flows in Figure \ref{figs:afunctionKasner}, computed as piecewise functions of $r$. $E$ is a stand-in for $\rho$ in the exterior $r < r_h$ and for $\kappa$ in the interior $r > r_h$. $a_T$ is stationary at the horizon $r = r_h$ because this is the IR fixed point.}
\label{figs:afunctionDerivKasner}
\end{figure}

The backreacted geometries are labeled by a dimensionless quantity called the ``deformation parameter." Explicitly this is $\phi_0/T^{d-\Delta}$, where $T$ is the black hole temperature \eqref{temp} and $\phi_0$ is the source of the scalar field read from the near-boundary ($r \to 0$) expansion
\begin{equation}
\phi(r) \sim \phi_0 r^{d-\Delta} + \frac{\expval{\mathcal{O}}}{2\Delta - d}r^\Delta, \qquad r \to 0.\label{bdryExp}
\end{equation}
Getting $\phi_0$ from $\phi(r)$ depends on the choice of quantization---whether we take $\Delta > d/2$ or $\Delta < d/2$ for a given $m$ \cite{Klebanov:1999tb}.\footnote{At $\Delta = d/2$ exactly, the near-boundary expansion \eqref{bdryExp} becomes functionally different and there are two ways to quantize \cite{Minces:1999eg}.} Here we use ``standard" quantization,
\begin{equation}
\Delta > \frac{d}{2} \implies \phi_0 = \lim_{r\to 0} r^{\Delta - d}\phi(r).
\end{equation}
Upon choosing $d$ and $\Delta$, one can plug the solutions of \eqref{eom1}--\eqref{eom3} into \eqref{aFunctionSch} for various deformation parameters. We do so in Figure \ref{figs:afunctionKasner} for $d = 3$, $\Delta = 2$. We also show how $a_T$ evolves along the entire flow in Figure \ref{figs:afunctionDerivKasner}.


\subsubsection{Probing the Singularity}

From our numerics, $a_T$ decreases monotonically along both the UV $\to$ IR and trans-IR segments of the free Kasner flow, validating the general monotonicity argument. The rate's dependence on $\phi_0/T$ appears nontrivial, but how much we can study this detailed analytic behavior is limited because our solutions are numerical.\footnote{\cite{Frenkel:2020ysx} found that, in $d = 3$, $\Delta = 2$, the near-singularity Kasner exponent $p_t$ has nontrivial dependence on $\phi_0/T$, achieving a maximum in the range $24 < \phi_0/T < 25$. It would be interesting to determine if this is related to the behavior in Figure \ref{figs:afunctionDerivKasner}.} 

Nonetheless, we can still address (Q2) in free Kasner flows analytically because we at least know the behavior of the metric functions near the singularity. We first combine the usual near-singularity logarithmic divergence of the scalar field (for some $q \in \mathbb{R}$) \cite{Doroshkevich:1978aq,Fournodavlos:2018lrk},
\begin{equation}
\phi(r) \sim (d-1)q\log r, \qquad r \to \infty,\label{asymptPhi}
\end{equation}
with the equations of motion \eqref{eom1}--\eqref{eom3}. Then \cite{Frenkel:2020ysx,Caceres:2021fuw},
\begin{equation}
\chi(r) \sim (d-1)q^2 \log r + \chi_1, \qquad r \to \infty,\label{asymptChi}
\end{equation}
where $\chi_1$ is another number. By plugging this into \eqref{aFunctionSch}, we have that for free Kasner flows,
\begin{equation}
a_T(r) \sim C_d r^{-(d-1)^2 q^2/2}, \qquad r \to \infty,\label{atKasner}
\end{equation}
where $C_d > 0$ is a constant. This expression goes to $0$ at the singularity for any $q > 0$,\footnote{$q = 0$ corresponds to AdS-Schwarzschild, i.e. no RG flow.} signalling a total loss of degrees of freedom. Because of monotonicity, the degrees of freedom ``thin out" as we approach the singularity.

We can also compute the near-singularity behavior of $da_T/d\kappa$ by noting for some positive number $f_1$,
\begin{equation}
F(r) \sim -f_1 r^{d + (d-1)q^2/2}, \qquad r \to \infty.\label{asymptF}
\end{equation}
Thus, we have that
\begin{equation}
\frac{da_T}{d\kappa} = r\sqrt{|F(r)|}\frac{da_T}{dr} \sim -\tilde{C}_d r^{\sigma},\ \ r \to \infty,
\end{equation}
where $\tilde{C}_d > 0$ is another constant and $\sigma$ is
\begin{equation}
\sigma = \frac{2d-q^2(d-1)(2d-3)}{4}.\label{expSing}
\end{equation}
This expression is more meaningful if it is written in terms of the Kasner exponent $p_t$, which is directly related to $q$ in the literature \cite{Frenkel:2020ysx,Caceres:2021fuw}. For any $d$,
\begin{align}
&q^2 = \frac{2[d(1+p_t) - 2]}{(d-1)(1-p_t)}\nonumber\\
&\implies \sigma = \frac{(d-1)[d(1+p_t) - 3]}{p_t - 1}.\label{sigma}
\end{align}
\textit{A priori}, there are three parametric regimes: $\sigma < 0$, $\sigma = 0$, and $\sigma > 0$. $\sigma$ has both a root at $p_t = -1 + 3/d \equiv p_t^*$ and a discontinuity at $p_t = 1$ at which
\begin{equation}
\lim_{p_t \to 1^-} \sigma = -\infty,\ \ \lim_{p_t \to 1^+} \sigma = +\infty.
\end{equation}
Additionally, $d\sigma/dp_t < 0$ for any $d \geq 2$. Thus,
\begin{align}
\begin{Bmatrix}
\sigma < 0\\
\sigma = 0\\
\sigma > 0
\end{Bmatrix} \iff \begin{Bmatrix}
p_t^* < p_t < 1\\
p_t = p_t^*\\
p_t < p_t^*\ \text{or}\ p_t > 1
\end{Bmatrix}.\label{sigmapt}
\end{align}
There are two ways to make use of \eqref{sigmapt}. One is to numerically compute the range of $p_t$ for particular free Kasner flows so as to constrain $\sigma$. \cite{Frenkel:2020ysx} does the first step for free flows with $d = 3$, $\Delta = 2$, and their numerics indicate that $\sigma > 0$ (implying that $da_T/d\kappa \to -\infty$). Thus, a numerical analysis can be used to examine the evolution of $a_T$ on a case-by-case basis.

The second approach is to take the negative blow-up of $da_T/d\kappa$ near the singularity as a given, thus insisting that $\sigma > 0$ for all free Kasner flows. This is more of a stretch, but if this holds then one would obtain analytic bounds on the near-singularity Kasner exponent $p_t$.

We reiterate that our analysis takes place in the near-singularity region, which is why $p_t$ has so much control over the evolution of $a_T$. This is powerful enough to address (Q2) since it concerns only the endpoint of the trans-IR, but one may ask about the rest of the trans-IR flow. Does $a_T$ feature any interesting dynamics aside from its monotonicity? We would expect such physics to be highly dependent upon the fields in the interior but infinitely far from the singularity.

\subsection{Considerations for Generalization}

The above discussion is focused on free Kasner flows, so we are motivated to extract whatever lessons we can for more general types of flows featuring both more intricate behavior along the trans-IR and more general singularity structure. We now make some broad statements regarding such generalizations, leaving further examination to future work. Based on these statements, we conjecture that $a_T$ vanishing at the singularity is a general feature of holographic trans-IR flows.

\subsubsection{General Kasner Singularities}

One may ask about Kasner flows sourced by more complicated matter sectors, such as the self-interacting $\phi^4$-scalar theory of \cite{Wang:2020nkd} or $\phi$ coupled to Maxwellian and/or axionic fields \cite{Hartnoll:2020fhc,Sword:2021pfm,Mansoori:2021wxf}. One could even consider dimensional reductions of supergravity theories \cite{Das:2021vjf} or modify the gravitational theory to include higher-curvature terms \cite{Grandi:2021ajl} (upon appropriate changes to $a_T$ \eqref{aFunction2} \cite{Myers:2010xs,Myers:2010tj}). Either scalar hair \cite{Hartnoll:2020rwq,Cai:2020wrp} or vector hair \cite{Cai:2021obq} generally prevents the formation of inner horizons in black holes, so trans-IR flows even in these theories will still end at a spacelike singularity, just as in the free scalar theory. Assuming a Kasner singularity, we may conjecture that $a_T$ still vanishes at the trans-IR endpoint.

There is evidence for our conjecture. Take the minimal holographic superconductor of \cite{Hartnoll:2020fhc} with a massive scalar field charged under a Maxwell field. The hairy black holes in this theory exhibit physics dependent upon the charge, such as Josephson oscillations of the scalar field and Kasner inversions---changes to the Kasner exponent due to instability. In fact, their numerics suggest the possibility of some flows\footnote{These occur at discrete values of the temperature (relative to the holographic superconductor's critical temperature).} with infinitely many Kasner inversions; these flows never ``settle down" and instead exhibit chaotic mixmaster behavior \cite{Misner:1969hg,Damour:2002et,Damour:2002tc}.

This rich geometrical structure suggests interesting dynamics for $a_T$ along the trans-IR flow. However, even in this case, $a_T$ decays with large $r$ \eqref{atKasner}. The power in this asymptotic expression for $a_T$ is always negative, so we would still expect $a_T \to 0$ in the strict limit $r \to \infty$. Assuming oscillations and inversions affect $a_T$, we expect them to emerge in the derivative $da_T/d\kappa$, which need not even be monotonic. One could numerically examine $\sigma$ in order to better understand the evolution of $a_T$.

\subsubsection{BKL Singularities}

The BKL program \cite{Lifshitz:1963ps,Belinskii:1970ew,Belinskii:1982pk} characterizes the near-singularity geometry allowed by the Einstein equations. Instead of obtaining numerical solutions for a particular matter sector, one can asymptotically analyze the equations of motion for classes of theories. We lose subtlety in $a_T$ away from the singularity in exchange for more near-singularity analytic power. To consider holographic RG flows, one would need matter fields dual to relevant deformations on the boundary and a negative cosmological constant (realizable as a constant potential term \cite{Damour:2002tc}).

The expected near-singularity geometry is either a Kasner universe or an infinite sequence of Kasner epochs undergoing mixmaster behavior, depending on the types of fields and the number of spacetime dimensions $D \equiv d+1$ \cite{Damour:2002et,Damour:2002tc}. For example, free Kasner flows have the former, and the flows of \cite{Hartnoll:2020fhc} with infinite Kasner inversions have the latter. If $a_T \to 0$ at spacelike singularities in both of these examples, it is natural to conjecture that $a_T \to 0$ at either kind of BKL singularity, too. One way forward may be to utilize the description of near-singularity dynamics as billiards in hyperbolic space \cite{Damour:2002et}.

\section{Probes of Trans-IR from Quantum Information}\label{sec:Q3}

While holographic RG flow is one perspective of the relationship between the bulk and the boundary, we may also take the view that bulk geometry encodes information-theoretic quantities about the boundary quantum states. By combining these two perspectives, such quantities can be thought of as probes of the trans-IR regime. Concretely, we may ask which quantities reach the singularity and thus probe the full trans-IR.

We examine several quantities for which holographic prescriptions have been well studied. For each one, we are concerned with the maximal radius $r_m$ that is reached by the dual bulk object---a measure of how far that particular information-theoretic quantity probes into the trans-IR. First, we argue that both entanglement entropy from the Ryu-Takayanagi (RT) and Hubeny-Rangamani-Takayanagi (HRT) prescriptions \cite{Ryu:2006bv,Hubeny:2007xt} and complexity from Complexity = Volume (CV) \cite{Susskind:2014rva,Stanford:2014jda,Susskind:2014moa} typically probe only some of the trans-IR; $r_m$ is bounded from above. Then, we find that both the 2-point correlator from geodesic approximation \cite{Balasubramanian:1999zv,Fidkowski:2003nf} and complexity from Complexity = Action (CA) \cite{Brown:2015bva,Brown:2015lvg} see the full trans-IR, with $r_m$ running over the full interior.

For static black holes, we generally have a ``critical" time $t_c > 0$ controlling which of the latter two quantities---the 2-point correlator or CA---at a particular boundary time $t_b$ is a ``good" probe of the trans-IR. This critical time has appeared before in work examining each probe individually \cite{Fidkowski:2003nf,Carmi:2017jqz}, but through the lens of holographic RG flow $t_c$ is a scale that characterizes the trans-IR flow. Concretely, noting that the maximal radius of either probe is a function of boundary time, a probe is good for the range of $t_b$ for which $r_m$ has continuous support. Given some critical time, we find that the 2-point correlator is the early-time ($|t_b| < t_c$) probe of the trans-IR while CA is the late-time ($|t_b| > t_c$) probe. This ``complementarity" between correlations and CA follows from a geodesic analysis of Schwarzschild-like black holes \eqref{schwarzLike}.

However, there are some caveats. First, the failure of entropy and CV to probe the full trans-IR mostly hold except for very specific, finely tuned situations. Specifically, there are particular choices of the number of spatial dimensions $d$ in which entanglement ($d = 2$) or CV ($d = 1$) may probe the full trans-IR. Additionally, there are also certain geometries (i.e. $d = 2$ AdS-Schwarzschild black holes) for which correlations and CA are not complementary; both are instead ``good" for all $t_b$. In fact, lack of complementarity becomes more common in charged or rotating black holes, as we discuss later.

Nonetheless, we reiterate that our focus is on flows represented by static black holes, in which case such exceptions require some sort of fine-tuning and so should not be drawn upon as general examples.

\subsection{Entanglement and CV Are Not Enough}\label{sec:spacelikeSurfBad}

We first make the case that both entanglement entropy and CV usually fail to fully probe the trans-IR. That is not to say that neither probes at least some of the trans-IR, nor are we asserting that they never probe the full trans-IR. Indeed, there are finely tuned situations for which the latter happens,\footnote{Entanglement entropy works in $d = 2$, while CV works in $d = 1$. This is specifically because they are encoded by geodesics.} but entanglement entropy and CV fail to do so in general. In the interest of brevity, we leave most of the technical details to Appendix \ref{app:B}.

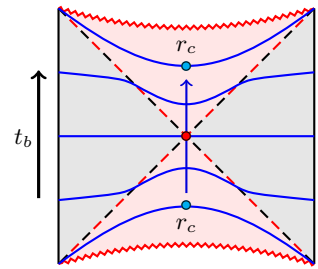
\begin{figure}
\centering
\begin{tikzpicture}
\draw[-,white] (0,0) to (0,3.4);
\draw[->,very thick] (0,0.9) to (0,2.6);
\node at (-0.2,1.7) {$t_b$};
\end{tikzpicture}\ \ 
\begin{tikzpicture}[scale=1.7]
\draw[-,draw=none,fill=black!10] (1,1) to (1,-1) to (0,0) to (1,1); 
\draw[-,draw=none,fill=black!10] (-1,1) to (-1,-1) to (0,0) to (-1,1);

\draw[-,draw=none,fill=red!10] (-1,1) ..controls (-0.3,0.8) and (0.3,0.8) .. (1,1) to (0,0) to (-1,1); 
\draw[-,draw=none,fill=red!10] (1,-1) ..controls (0.3,-0.8) and (-0.3,-0.8) .. (-1,-1) to (0,0) to (1,-1);

\draw[-,thick,red,decoration = {zigzag,segment length = 1mm, amplitude = 0.25mm},decorate] (-1,1) ..controls (-0.3,0.8) and (0.3,0.8) .. (1,1);
\draw[-,thick] (1,1) to (1,-1);
\draw[-,thick,red,decoration = {zigzag,segment length = 1mm, amplitude = 0.25mm},decorate] (1,-1) ..controls (0.3,-0.8) and (-0.3,-0.8) .. (-1,-1);
\draw[-,thick] (-1,-1) to (-1,1);

\draw[-,dashed,thick,dash pattern= on 4pt off 8pt,dash phase=6pt,red] (-1,1) to (1,-1);
\draw[-,dashed,thick,dash pattern= on 4pt off 8pt,dash phase=6pt,red] (1,1) to (-1,-1);

\draw[-,dashed,thick,dash pattern= on 4pt off 8pt] (-1,1) to (1,-1);
\draw[-,dashed,thick,dash pattern= on 4pt off 8pt] (1,1) to (-1,-1);

\draw[-,blue,thick] (-1,-1) .. controls (-0.2,-0.4) and (0.2,-0.4) .. (1,-1);
\draw[-,blue,thick] (-1,-0.5) .. controls (-0.5,-0.45) .. (-0.4,-0.4) .. controls (-0.05,-0.2) and (0.05,-0.2) .. (0.4,-0.4) .. controls (0.5,-0.45) .. (1,-0.5);
\draw[-,blue,thick] (-1,0) to (1,0);
\draw[-,blue,thick] (-1,0.5) .. controls (-0.5,0.45) .. (-0.4,0.4) .. controls (-0.05,0.2) and (0.05,0.2) .. (0.4,0.4) .. controls (0.5,0.45) .. (1,0.5);
\draw[-,blue,thick] (-1,1) .. controls (-0.2,0.4) and (0.2,0.4) .. (1,1);

\draw[->,black!15!blue,thick] (0,-0.445) to (0,0.445);

\node[cyan] at (0,0.545) {$\bullet$};
\node at (0,0.545) {$\circ$};
\node at (0,0.7) {$r_c$};

\node[cyan] at (0,-0.545) {$\bullet$};
\node at (0,-0.545) {$\circ$};
\node at (0,-0.7) {$r_c$};

\node[red] at (0,0) {$\bullet$};
\node at (0,0) {$\circ$};
\end{tikzpicture}\ \ 
\begin{tikzpicture}
\draw[-,white] (0,0) to (0,3.4);
\draw[white,->,very thick] (0,0.9) to (0,2.6);
\node[white] at (0.2,1.7) {$t_b$};
\end{tikzpicture}
\caption{Extremal spacelike surfaces of dimension $\geq 2$ which symmetrically connect the disjoint boundaries of a two-sided, asymptotically AdS black hole at various boundary times $t_b$. The maximal radius $r_m$ as a function of $t_b$ takes on values in the interval $[r_h,r_c)$, approaching $r_c$ as $|t_b| \to \infty$.}
\label{figs:extremalSpacelikeSurfaces}
\end{figure}

Fundamentally, this failure is because of the existence of ``barriers" preventing the pertinent boundary-anchored extremal spacelike surfaces from continuously (with respect to the $t_b = 0$ slice) reaching the singularity (Figure \ref{figs:extremalSpacelikeSurfaces}). Such barriers have been seen explicitly in both AdS-Schwarzschild \cite{Hartman:2013qma} and free Kasner flows \cite{Frenkel:2020ysx,Caceres:2021fuw} (reviewed in Appendix \ref{app:B1}) for particular surfaces but are supposed to be rather general features of asymptotically locally AdS spacetimes \cite{Wall:2012uf,Engelhardt:2013tra}. For these extremal spacelike surfaces, $r_m$ is bounded from above by a finite radius $r_c$, so there is an upper limit for how far into the trans-IR they can probe.

We start with entanglement entropy, which is known to be a monotonic function in both holographic \cite{Myers:2010xs,Myers:2010tj} and nonholographic \cite{Casini:2016udt,Casini:2017vbe} RG flow. Recall the RT presecription \cite{Ryu:2006bv} and its covariant extension \cite{Hubeny:2007xt}: for a boundary CFT subregion $\mathcal{R}$, its entanglement entropy $S(\mathcal{R})$ is calculated to leading order in $1/G_N$ by minimizing the area functional on codimension-2 extremal surfaces $\Sigma$ homologous to $\mathcal{R}$,
\begin{equation}
S(\mathcal{R}) = \substack{\text{\normalsize{min\,ext}}\\{\Sigma \sim \mathcal{R}}} \frac{A[\Sigma]}{4G_N}.
\end{equation}
We consider $\mathcal{R}$ at a particular boundary time $t_b$ and denote the corresponding minimal-area surface as $\Sigma_0$. If $\mathcal{R}$ is a Cauchy slice of one of the disjoint boundaries, then $\Sigma_0$ is just the black hole horizon. Furthermore by entanglement wedge nesting \cite{Akers:2016ugt,Akers:2017ttv}, if $\mathcal{R}$ is a subregion of this Cauchy slice then $\Sigma_0$ is entirely in the black hole exterior.

To get a $\Sigma_0$ which probes the interior, $\mathcal{R}$ must include intervals on both sides. Fixing a constant $x_\mathcal{R} \in \mathbb{R}$, the symmetrical case for which we take $\mathcal{R}$ to be
\begin{equation}
t = t_b,\ \ x^1 \in [x_\mathcal{R},\infty),
\end{equation}
yields a Hartman-Maldacena (HM) surface \cite{Hartman:2013qma}. These are the surfaces obtained in Appendix \ref{app:B} for $k = 1$. However, these still fail to probe the full trans-IR regime unless $d = 2$ (in which case HM surfaces are geodesics).

We now discuss CV \cite{Susskind:2014rva,Stanford:2014jda,Susskind:2014moa}, in which the boundary-time-dependent complexity of the UV state $\mathcal{C}_V(t_b)$ is identified as the volume of a maximal, codimension-1 bulk slice $V(t_b)$,
\begin{equation}
\mathcal{C}_V(t_b) = \frac{8\pi}{\ell_P^{d-1}} V(t_b).
\end{equation}
Such volumes are straightforward to write as integrals of metric functions when considering spherically symmetric metrics such as \eqref{schwarzLike} \cite{Carmi:2017jqz,Hashimoto:2021umd}. Indeed, the volume is explicitly written in \eqref{volTot} by setting $k = 0$.

Just like HM surfaces, this bulk slice will fail to probe the full trans-IR (apart from if $d = 1$, when these slices are simply geodesics). In general, just like the HM surfaces, $r_m$ will be bounded and the slices will get stuck infinitely far from the singularity.

\subsection{Complementarity of Correlations and CA}\label{sec:complementarity}

While entanglement and CV are not enough to probe the full trans-IR flow, there are two quantities describing quantum information of the boundary state which do. These are the (secondary sheet of the \cite{Fidkowski:2003nf}) 2-point correlator and CA.

Assume flows corresponding to static black holes. In this case, we will see that the 2-point correlator and CA generally probe the trans-IR in a complementary manner dictated by a flow-dependent critical time,
\begin{equation}
t_c = P\int_0^\infty \frac{e^{\chi(r)/2}}{F(r)} dr,\label{formulatc}
\end{equation}
so long as $t_c > 0$.

To see why this happens, without loss of generality, consider $t_b \geq 0$. In the early-time ($0 \leq t_b < t_c$) part of the UV state, the 2-point correlator probes the full trans-IR. Any symmetric spacelike geodesic dual to this correlator at a particular boundary time will have a maximal radius $r_m$ between the horizon and the singularity. However, as $t_b \to t_c$ from below, $r_m \to \infty$ and the geodesic becomes ``nearly" null---approaching the appearance of two symmetric null rays fired from opposite connected components of the boundary.

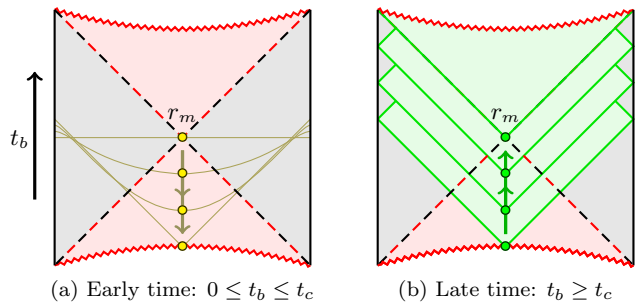
\begin{figure}
\centering
\begin{tikzpicture}
\draw[-,white] (0,0) to (0,3.4);
\draw[->,very thick] (0,0.9) to (0,2.6);
\node at (-0.2,1.7) {$t_b$};
\end{tikzpicture}
\subfloat[Early time: $0 \leq t_b \leq t_c$\label{figs:early}]
{
\begin{tikzpicture}[scale=1.7]
\draw[-,draw=none,fill=black!10] (1,1) to (1,-1) to (0,0) to (1,1); 
\draw[-,draw=none,fill=black!10] (-1,1) to (-1,-1) to (0,0) to (-1,1);

\draw[-,draw=none,fill=red!10] (-1,1) ..controls (-0.3,0.8) and (0.3,0.8) .. (1,1) to (0,0) to (-1,1); 
\draw[-,draw=none,fill=red!10] (1,-1) ..controls (0.3,-0.8) and (-0.3,-0.8) .. (-1,-1) to (0,0) to (1,-1);

\draw[-,black!40!yellow] (-1,0.145) to (0,-0.855) to (1,0.145);

\draw[-,black!40!yellow] (-1,0.145*2/3) .. controls (-0.9,0.145*2/3) and (-0.45,-0.575) .. (0,-0.855+0.285) .. controls (0.45,-0.575) and (0.9,0.145*2/3) .. (1,0.145*2/3);

\draw[-,black!40!yellow] (-1,0.145/3) .. controls (-0.875,0.145/3) and (-0.65,-0.25) .. (0,-0.855+0.57) .. controls (0.65,-0.25) and (0.875,0.145/3) .. (1,0.145/3);

\draw[-,black!40!yellow] (-1,0) to (1,0);

\draw[-,thick,red,decoration = {zigzag,segment length = 1mm, amplitude = 0.25mm},decorate] (-1,1) ..controls (-0.3,0.8) and (0.3,0.8) .. (1,1);
\draw[-,thick] (1,1) to (1,-1);
\draw[-,thick,red,decoration = {zigzag,segment length = 1mm, amplitude = 0.25mm},decorate] (1,-1) ..controls (0.3,-0.8) and (-0.3,-0.8) .. (-1,-1);
\draw[-,thick] (-1,-1) to (-1,1);

\draw[-,dashed,thick,dash pattern= on 4pt off 8pt,dash phase=6pt,red] (-1,1) to (1,-1);
\draw[-,dashed,thick,dash pattern= on 4pt off 8pt,dash phase=6pt,red] (1,1) to (-1,-1);

\draw[-,dashed,thick,dash pattern= on 4pt off 8pt] (-1,1) to (1,-1);
\draw[-,dashed,thick,dash pattern= on 4pt off 8pt] (1,1) to (-1,-1);

\draw[<-,very thick,black!50!yellow] (0,-0.4675) to (0,-0.1);
\draw[<-,very thick,black!50!yellow] (0,-0.755) to (0,-0.1);

\node[yellow] at (0,-0.855) {$\bullet$};
\node at (0,-0.855) {$\circ$};

\node[yellow] at (0,-0.855+0.285) {$\bullet$};
\node at (0,-0.855+0.285) {$\circ$};

\node[yellow] at (0,-0.855+0.57) {$\bullet$};
\node at (0,-0.855+0.57) {$\circ$};

\node[yellow] at (0,0) {$\bullet$};
\node at (0,0) {$\circ$};

\node at (0,0.175) {$r_m$};
\end{tikzpicture}
}\qquad
\subfloat[Late time: $t_b \geq t_c$\label{figs:late}]
{
\begin{tikzpicture}[scale=1.7]
\draw[-,draw=none,fill=black!10] (1,1) to (1,-1) to (0,0) to (1,1); 
\draw[-,draw=none,fill=black!10] (-1,1) to (-1,-1) to (0,0) to (-1,1);

\draw[-,draw=none,fill=red!10] (-1,1) ..controls (-0.3,0.8) and (0.3,0.8) .. (1,1) to (0,0) to (-1,1); 
\draw[-,draw=none,fill=red!10] (1,-1) ..controls (0.3,-0.8) and (-0.3,-0.8) .. (-1,-1) to (0,0) to (1,-1);

\draw[-,thick,black!10!green,fill=black!10!green!10] (-1,0.145) to (0,-0.855) to (1,0.145) to (0+0.2,1.145-0.2) to (0-0.2,1.145-0.2) to (-1,0.145);

\draw[-,thick,black!10!green,fill=black!10!green!10] (-1,0.145+0.285) to (0,-0.855+0.285) to (1,0.145+0.285) to (0+0.485,1.145-0.485+0.285) to (0-0.485,1.145-0.485+0.285) to (-1,0.145+0.285);

\draw[-,thick,black!10!green,fill=black!10!green!10] (-1,0.145+0.57) to (0,-0.855+0.57) to (1,0.145+0.57) to (0+0.77,1.145-0.77+0.57) to (0-0.77,1.145-0.77+0.57) to (-1,0.145+0.57);

\draw[-,thick,draw=none,fill=black!10!green!10] (1-0.02,1-0.005) to (0,0.865) to (-1+0.02,1-0.005) to (0,0) to (1,1);

\draw[-,draw=none,fill=white] (-1,1) ..controls (-0.3,0.8) and (0.3,0.8) .. (1,1) to (-1,1);

\draw[-,dashed,thick,dash pattern= on 4pt off 8pt,dash phase=6pt,red] (-1,1) to (1,-1);
\draw[-,dashed,thick,dash pattern= on 4pt off 8pt,dash phase=6pt,red] (1,1) to (-1,-1);

\draw[-,dashed,thick,dash pattern= on 4pt off 8pt] (-1,1) to (1,-1);
\draw[-,dashed,thick,dash pattern= on 4pt off 8pt] (1,1) to (-1,-1);

\draw[-,thick,black!10!green] (1,1) to (0,0) to (-1,1);

\draw[-,thick,red,decoration = {zigzag,segment length = 1mm, amplitude = 0.25mm},decorate,fill=white] (-1,1) ..controls (-0.3,0.8) and (0.3,0.8) .. (1,1);
\draw[-,thick] (1,1) to (1,-1);
\draw[-,thick,red,decoration = {zigzag,segment length = 1mm, amplitude = 0.25mm},decorate,fill=white] (1,-1) ..controls (0.3,-0.8) and (-0.3,-0.8) .. (-1,-1);
\draw[-,thick] (-1,-1) to (-1,1);

\draw[-,thick,red,decoration = {zigzag,segment length = 1mm, amplitude = 0.25mm},decorate] (1,-1) ..controls (0.3,-0.8) and (-0.3,-0.8) .. (-1,-1);

\draw[->,very thick,black!30!green] (0,-0.755) to (0,-0.375);
\draw[->,very thick,black!30!green] (0,-0.755) to (0,-0.1);

\node[green] at (0,0) {$\bullet$};
\node at (0,0) {$\circ$};

\node[green] at (0,-0.855+0.285) {$\bullet$};
\node at (0,-0.855+0.285) {$\circ$};

\node[green] at (0,-0.855+0.57) {$\bullet$};
\node at (0,-0.855+0.57) {$\circ$};

\node[green] at (0,-0.855) {$\bullet$};
\node at (0,-0.855) {$\circ$};

\node at (0,0.175) {$r_m$};
\end{tikzpicture}
}
\caption{The complementarity of the 2-point correlator (a) and holographic complexity from CA (b) for $t_b \geq 0$. From the figure, it is evident that the null limit of the symmetric spacelike geodesics coincides precisely with the past-directed null sheet bounding the WDW patch at $t_b = t_c$.}
\label{figs:maxComplementarity}
\end{figure}

Precisely at that moment, such null geodesics constitute the past-directed null sheet bounding the Wheeler-DeWitt (WDW) patch at $t_b = t_c$, with their intersection being a ``joint" located at the singularity. This particular joint moves towards the horizon as $t_b \to \infty$; taking its position as the maximal radius $r_m$ for CA, $r_m \to r_h$ for the WDW patches encoding late-time ($t_b > t_c$) complexity of the UV state. The point is that the UV state sees the full trans-IR flow through different probes for complementary intervals of $t_b$:
\begin{align}
|t_b| < t_c &\iff \text{2-point correlator},\\
|t_b| > t_c &\iff \text{Complexity = Action}.
\end{align}
See Figure \ref{figs:maxComplementarity} for a visual representation of this phenomenon.

However, note that this complementarity is not the only possibility, since $t_c$ need not be positive \textit{a priori}.\footnote{Complementarity will turn out to be related to the shape of the black hole, which can be diagnosed by a formula for the critical time $t_c$. Specifically, complementarity is a feature of black holes with $t_c > 0$, which when drawn with vertical boundaries have future singularities bending ``down" \cite{Auzzi:2022bfd}.} Even in the static case, there are specific finely tuned black holes for which $t_c = 0$ where complementarity is not a feature---we instead have a ``simultaneity" of the probes because the symmetric spacelike geodesics never become null. More generally, black holes with angular momentum or charge may not feature complementarity for larger regions of parameter space consisting of black holes with $t_c < 0$. We briefly discuss these cases in Section \ref{sec:noComp}. Nonetheless, there we argue that $t_c > 0$ is a general condition for the static case, and our numerics for free Kasner flows (Appendix \ref{app:C}) serve as evidence.


\subsubsection{Early Times: 2-point Correlator}\label{sec:geoApprox}

In AdS/CFT, correlation functions of the boundary are encoded by bulk paths connecting the insertion points \cite{Balasubramanian:1999zv}. Specifically, for a scalar operator $\widetilde{\mathcal{O}}$ with conformal dimension $\widetilde{\Delta}$, the 2-point correlator is a path integral,
\begin{equation}
\expval{\widetilde{\mathcal{O}}(\zeta_1)\widetilde{\mathcal{O}}(\zeta_2)} = \int_{\zeta_1 \to \zeta_2} \mathcal{D}\mathcal{P}\,e^{-\widetilde{\Delta} L[\mathcal{P}]},\label{pathIntGeo}
\end{equation}
where $\mathcal{P}$ is a bulk path from $\zeta_1$ to $\zeta_2$ while $L[\mathcal{P}]$ is the renormalized\footnote{We may renormalize by either using a cutoff surface or performing background subtraction. We do not specify a renormalization scheme here.} length of $\mathcal{P}$. In the ``heavy" limit $\widetilde{\Delta} \to \infty$, $\widetilde{\mathcal{O}}$ is irrelevant and \eqref{pathIntGeo} becomes a sum over saddles of the length functional, i.e. a sum over geodesics,
\begin{equation}
\expval{\widetilde{\mathcal{O}}(\zeta_1)\widetilde{\mathcal{O}}(\zeta_2)} \sim \sum_{\text{geodesics}} e^{-\widetilde{\Delta} L(\zeta_1,\zeta_2)},\ \ \widetilde{\Delta} \to \infty.
\end{equation}
We are concerned with the 2-point correlator between symmetric insertions on disjoint boundaries of asymptotically AdS black holes. Thus the pertinent bulk objects are symmetric spacelike geodesics.\footnote{\cite{Fidkowski:2003nf} notes that such geodesics actually encode a branch of the analytic continuation of the boundary theory's correlator. We will address this point later, but for now we allow ourselves an abuse of terminology in referring to this branch as ``the correlator."} They are described by the expressions in Appendix \ref{app:B} with $k = d-1$, so they each have a characteristic energy $\mathcal{E}$ corresponding to some boundary time $t_b$ and maximal radius $r_m$.

\begin{figure}
\centering
\begin{tikzpicture}
\draw[-,white] (0,0) to (0,3.4);
\draw[->,very thick] (0,0.9) to (0,2.6);
\node at (-0.2,1.7) {$t_b$};
\end{tikzpicture}
\subfloat[General ($t_c > 0$)\label{figs:generalGeo}]
{
\begin{tikzpicture}[scale=1.7]
\draw[-,draw=none,fill=black!10] (1,1) to (1,-1) to (0,0) to (1,1); 
\draw[-,draw=none,fill=black!10] (-1,1) to (-1,-1) to (0,0) to (-1,1);

\draw[-,draw=none,fill=red!10] (-1,1) ..controls (-0.3,0.8) and (0.3,0.8) .. (1,1) to (0,0) to (-1,1); 
\draw[-,draw=none,fill=red!10] (1,-1) ..controls (0.3,-0.8) and (-0.3,-0.8) .. (-1,-1) to (0,0) to (1,-1);

\draw[-,black!40!yellow] (-1,0.145) to (0,-0.855) to (1,0.145);

\draw[-,black!40!yellow] (-1,0.145*2/3) .. controls (-0.9,0.145*2/3) and (-0.45,-0.575) .. (0,-0.855+0.285) .. controls (0.45,-0.575) and (0.9,0.145*2/3) .. (1,0.145*2/3);

\draw[-,black!40!yellow] (-1,0.145/3) .. controls (-0.875,0.145/3) and (-0.65,-0.25) .. (0,-0.855+0.57) .. controls (0.65,-0.25) and (0.875,0.145/3) .. (1,0.145/3);

\draw[-,black!40!yellow] (-1,0) to (1,0);

\draw[-,black!40!yellow] (-1,-0.145/3) .. controls (-0.875,-0.145/3) and (-0.65,0.25) .. (0,0.855-0.57) .. controls (0.65,0.25) and (0.875,-0.145/3) .. (1,-0.145/3);

\draw[-,black!40!yellow] (-1,-0.145*2/3) .. controls (-0.9,-0.145*2/3) and (-0.45,0.575) .. (0,0.855-0.285) .. controls (0.45,0.575) and (0.9,-0.145*2/3) .. (1,-0.145*2/3);

\draw[-,black!40!yellow] (-1,-0.145) to (0,0.855) to (1,-0.145);

\draw[-,thick,red,decoration = {zigzag,segment length = 1mm, amplitude = 0.25mm},decorate] (-1,1) ..controls (-0.3,0.8) and (0.3,0.8) .. (1,1);
\draw[-,thick] (1,1) to (1,-1);
\draw[-,thick,red,decoration = {zigzag,segment length = 1mm, amplitude = 0.25mm},decorate] (1,-1) ..controls (0.3,-0.8) and (-0.3,-0.8) .. (-1,-1);
\draw[-,thick] (-1,-1) to (-1,1);

\draw[-,dashed,thick,dash pattern= on 4pt off 8pt,dash phase=6pt,red] (-1,1) to (1,-1);
\draw[-,dashed,thick,dash pattern= on 4pt off 8pt,dash phase=6pt,red] (1,1) to (-1,-1);

\draw[-,dashed,thick,dash pattern= on 4pt off 8pt] (-1,1) to (1,-1);
\draw[-,dashed,thick,dash pattern= on 4pt off 8pt] (1,1) to (-1,-1);

\node[red] at (0,0) {$\bullet$};
\node at (0,0) {$\circ$};
\end{tikzpicture}
}\qquad
\subfloat[Finely tuned ($t_c = 0$)\label{figs:geoFine}]
{
\begin{tikzpicture}[scale=1.7]
\draw[-,draw=none,fill=black!10] (1,1) to (1,-1) to (0,0) to (1,1); 
\draw[-,draw=none,fill=black!10] (-1,1) to (-1,-1) to (0,0) to (-1,1);

\draw[-,draw=none,fill=red!10] (-1,1) ..controls (-0.3,0.8) and (0.3,0.8) .. (1,1) to (0,0) to (-1,1); 
\draw[-,draw=none,fill=red!10] (1,-1) ..controls (0.3,-0.8) and (-0.3,-0.8) .. (-1,-1) to (0,0) to (1,-1);

\draw[-,black!40!yellow] (-1,-0.9) .. controls (-0.8,-0.9) and (-0.3,-0.75) .. (0,-0.75) .. controls (0.3,-0.75) and (0.8,-0.9) .. (1,-0.9);

\draw[-,black!40!yellow] (-1,-0.6) .. controls (-0.6,-0.6) and (-0.3,-0.4) .. (0,-0.4) .. controls (0.3,-0.4) and (0.6,-0.6) .. (1,-0.6);

\draw[-,black!40!yellow] (-1,-0.3) .. controls (-0.6,-0.3) and (-0.3,-0.2) .. (0,-0.2) .. controls (0.3,-0.2) and (0.6,-0.3) .. (1,-0.3);

\draw[-,black!40!yellow] (-1,0) to (1,0);

\draw[-,black!40!yellow] (-1,0.3) .. controls (-0.6,0.3) and (-0.3,0.2) .. (0,0.2) .. controls (0.3,0.2) and (0.6,0.3) .. (1,0.3);

\draw[-,black!40!yellow] (-1,0.6) .. controls (-0.6,0.6) and (-0.3,0.4) .. (0,0.4) .. controls (0.3,0.4) and (0.6,0.6) .. (1,0.6);

\draw[-,black!40!yellow] (-1,0.9) .. controls (-0.8,0.9) and (-0.3,0.75) .. (0,0.75) .. controls (0.3,0.75) and (0.8,0.9) .. (1,0.9);

\draw[-,thick,red,decoration = {zigzag,segment length = 1mm, amplitude = 0.25mm},decorate] (-1,1) ..controls (-0.3,0.8) and (0.3,0.8) .. (1,1);
\draw[-,thick] (1,1) to (1,-1);
\draw[-,thick,red,decoration = {zigzag,segment length = 1mm, amplitude = 0.25mm},decorate] (1,-1) ..controls (0.3,-0.8) and (-0.3,-0.8) .. (-1,-1);
\draw[-,thick] (-1,-1) to (-1,1);

\draw[-,dashed,thick,dash pattern= on 4pt off 8pt,dash phase=6pt,red] (-1,1) to (1,-1);
\draw[-,dashed,thick,dash pattern= on 4pt off 8pt,dash phase=6pt,red] (1,1) to (-1,-1);

\draw[-,dashed,thick,dash pattern= on 4pt off 8pt] (-1,1) to (1,-1);
\draw[-,dashed,thick,dash pattern= on 4pt off 8pt] (1,1) to (-1,-1);

\node[red] at (0,0) {$\bullet$};
\node at (0,0) {$\circ$};
\end{tikzpicture}
}
\caption{The symmetric spacelike geodesics in asymptotically AdS static black holes which are homotopic to the geodesics anchored at $t_b = 0$. Generally, the geodesics take the form shown in (a), becoming nearly null and thus producing a light-cone singularity. In specific finely tuned cases however, the geodesics may appear as in (b) in which case there is no light-cone singularity. For the purposes of our discussion, we will focus more on the general cases.}
\label{figs:spacelikeGeodesics}
\end{figure}
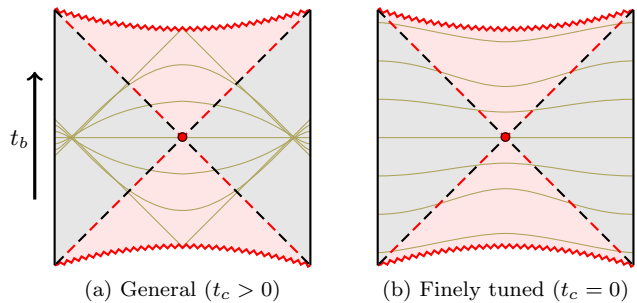

To see how the correlator evolves in time, we consider how the geodesics evolve with respect to $t_b$ (or, equivalently, $\mathcal{E}$). Starting at $t_b = 0$, regardless of the geometrical details, the geodesic resides entirely on the $t = 0$ bulk Cauchy slice and has energy $\mathcal{E} = 0$. \textit{A priori} there are two possibilities:
\begin{itemize}
\item We can take the limit $|\mathcal{E}| \to \infty$. \eqref{constraint1} implies that $r_m \to \infty$ (by regularity of the metric functions) while \eqref{constraint2} implies that $t_b$ goes to the aforementioned critical time,
\begin{equation}
|t_b| \to P\int_0^{\infty} \frac{e^{\chi(r)/2}}{F(r)} dr = t_c > 0.\label{critTimeGeo}
\end{equation}
\item $|\mathcal{E}|$ is bounded from above. Nonetheless we may still take $r_m \to \infty$. In the integrand of \eqref{constraint2}, we cannot suppress the pole as $r_m \to \infty$ because we cannot first take $|\mathcal{E}| \to \infty$. Thus, $|t_b| \to \infty$.
\end{itemize}
These two cases are illustrated in Figure \ref{figs:spacelikeGeodesics}. Figure \ref{figs:generalGeo} is the first case in which taking $|\mathcal{E}| \to \infty$ produces the nearly null geodesics. Figure \ref{figs:geoFine} is the second case where we simply have spacelike geodesics for all $t_b$.

There is a simple way to test which case we have---by assuming the first case and explicitly computing the integral in \eqref{critTimeGeo}. We will always get $t_c > 0$ for the first case. However, if the integral yields $t_c = 0$, our assumption must be wrong since we would get that $t_b = 0$ for both $\mathcal{E} = 0$ and $|\mathcal{E}| = \infty$. Thus, $|\mathcal{E}|$ must be bounded and we are left with the second case. We call the first case ``general" and the second case ``finely tuned," since the geometry must be chosen such that $t_c = 0$ for the latter.

For example, consider the planar AdS-Schwarzschild black holes discussed by \cite{Fidkowski:2003nf} and for which we compute $t_c$ in Appendix \ref{app:C} as \eqref{adsSchwarzCrit}. The critical time is $0$ if and only if $d = 2$. Additionally, plugging the appropriate metric functions ($\chi(r) = 0$, $F(r) = 1-(r/r_h)^d$) into \eqref{constraint1} yields
\begin{equation}
\mathcal{E}^2 = \frac{r_m^{d-2}}{r_h^d} - \frac{1}{r_m^2}.
\end{equation}
So $|\mathcal{E}|$ monotonically goes from $0$ to $1/r_h$ as $r_m \to \infty$ if $d = 2$, whereas $d > 2$ implies that $|\mathcal{E}|$ monotonically goes from $0$ to $\infty$ as $r_m \to \infty$. Setting $d = 2$ is the fine-tuning needed to get Figure \ref{figs:geoFine}.

So far, we have that $t_c > 0$ and the corresponding nearly null geodesics shown in Figure \ref{figs:generalGeo} feature in general black holes. Furthermore, $r_m(t_b)$ in these geometries starts at $r_h$ (for $|t_b| = 0$) and goes to $\infty$ (as $|t_b| \to t_c$). So, the 2-point correlator encoded by these geodesics only probes the trans-IR flow for $|t_b| < t_c$. \cite{Fidkowski:2003nf} argues that this correlator has a ``light-cone" singularity,\footnote{\cite{Fidkowski:2003nf} focuses on AdS-Schwarzschild, but this singularity is also seen in Kasner flows by \cite{Frenkel:2020ysx}. See also \cite{Festuccia:2005pi}.}
\begin{equation}
\expval{\widetilde{\mathcal{O}}_L(t_b)\widetilde{\mathcal{O}}_R(t_b)} \sim \begin{cases}\dfrac{1}{|t_b - t_c|^{2\widetilde{\Delta}}}&\text{if}\ t_b \to t_c,\\
\dfrac{1}{|t_b + t_c|^{2\widetilde{\Delta}}}&\text{if}\ t_b \to -t_c.
\end{cases}
\end{equation}
$\widetilde{\mathcal{O}}_{L/R}(t_b)$ is the value of the operator $\widetilde{\mathcal{O}}$ on the left/right boundary at $t_b$. However, this light-cone singularity is not expected from general properties of the boundary theory. The resolution of \cite{Fidkowski:2003nf} is that the expected boundary-state correlator is realized as a ``complexified" geodesic (as in $\mathcal{E} \in \mathbb{C}$), and that the light-cone singularity merely occurs on a secondary sheet of the analytic continuation of that correlator. Thus, our precise statement is that it is the secondary sheet of the 2-point correlator which probes the trans-IR regime for $|t_b| < t_c$.

\subsubsection{Late Times: Complexity = Action}

\begin{figure}
\centering
\begin{tikzpicture}[scale=0.9]
\draw[-,fill=black!10] (0,-0.5) to (0,3) to (1,4) to (1,0.5) to (0,-0.5);

\draw[-,draw=none,fill=black!10!green!10] (0,1.5) to (1,2.5) to (3,4) to (3,0.5) to (2,-0.5) to (0,1.5);
\draw[-,draw=none, fill=black!10!green!30] (0,1.5) to (1,2.5) to (3,4) to (2,3) to (0,1.5);
\draw[-,draw=none,fill=black!10!green!30] (0,1.5) to (1,2.5) to (3,0.5) to (2,-0.5) to (0,1.5);

\draw[-,thick,black!10!green] (0,1.5) to (1,2.5) to (3,4);
\draw[-,thick,black!10!green] (2,3) to (0,1.5);
\draw[-,thick,black!10!green,dashed] (3,4) to (2,3);

\draw[-,thick,black!10!green] (0,1.5) to (1,2.5) to (3,0.5);
\draw[-,thick,black!10!green] (2,-0.5) to (0,1.5);
\draw[-,thick,black!10!green,dashed] (3,0.5) to (2,-0.5);

\draw[-,dashed,thick,black!10!green] (3,4) to (3,0.5);
\draw[-,dashed,thick,black!10!green] (2,3) to (2,-0.5);

\node at (0.5,2.35) {$t_b$};
\node at (2.5,1.75) {$\mathcal{W}(t_b)$};

\end{tikzpicture}
\caption{A cartoon depiction of the Wheeler-DeWitt patch $\mathcal{W}(t_b)$ (the green solid wedge) obtained by shooting null rays from a constant-$t_b$ boundary slice into the bulk.}
\label{figs:wdwpatch}
\end{figure}
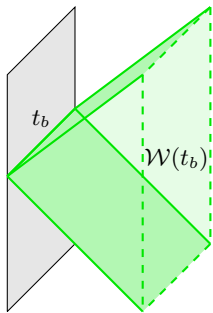

We now consider CA \cite{Brown:2015bva,Brown:2015lvg}, in which we associate the boundary-time-dependent complexity of the UV state $\mathcal{C}_A(t_b)$ with the action evaluated on the corresponding WDW patch $\mathcal{W}(t_b)$. This WDW patch is properly defined as the union of all bulk spatial slices anchored to the $t_b$ slice of the boundary, so it is found by shooting both past-directed and future-directed null rays into the bulk (Figure \ref{figs:wdwpatch}). Concretely, we write
\begin{equation}
\mathcal{C}_A(t_b) = \frac{I[\mathcal{W}(t_b)]}{\pi\hbar},
\end{equation}
where $I[\mathcal{W}(t_b)]$ is the bulk action on $\mathcal{W}(t_b)$.

CA is a rather unwieldy prescription compared to CV. On top of integrating the bulk action on the codimension-0 interior of the WDW patch, we must also account for the codimension-1 boundary terms and, notably for our purposes, the codimension-2 joint terms, with careful attention towards the null parts of the geometry \cite{Lehner:2016vdi}. Thus, $I[\mathcal{W}(t_b)]$ truly consists of three types of contributions,\footnote{This is the formal prescription, but generally the action diverges. This can be sidestepped by introducing regulator surfaces, but this also introduces additional boundary and joint terms.}
\begin{equation}
I[\mathcal{W}(t_b)] = I_{\mathcal{W}} + I_{\mathcal{B}} + I_{\mathcal{J}},\label{complexityWDW}
\end{equation}
which are respectively the bulk WDW, boundary, and joint terms.

For two-sided black holes, we can think of the WDW patch corresponding to a particular $t_b$ as being formed by shooting symmetric null rays from both asymptotic regions. However, the null rays can reach the singularity without intersecting one another. We use the null rays discussed in Appendix \ref{app:A} to note that there exists a branch-independent\footnote{This refers to the $\gamma$ discussed in Appendix \ref{app:A}.} critical time $t_c > 0$,
\begin{equation}
t_c = P\int_0^\infty \frac{e^{\chi(r)/2}}{F(r)} dr,
\end{equation}
for which there is a ``null-to-null" joint (whose position is taken as the maximal radius $r_m$) if and only if $|t_b| \geq t_c$. Specifically, for $t_b \leq -t_c$, the joint formed by future-directed null rays goes from the bifurcation point ($r_m = r_h$) at $t_b = -\infty$ to the future singularity ($r_m = \infty$) as $t_b \to -t_c$. If $|t_b| < t_c$, there is no joint. Then, as $t_b \to t_c$, a joint formed by past-directed null rays emerges at the past singularity ($r_m = \infty$), subsequently moving towards the bifurcation point again as $t_b \to \infty$.

So long as $t_c > 0$---the general case from before---the presence of a joint occurs in a complementary range of boundary time to the interval for which we have the symmetric spacelike geodesics of Section \ref{sec:geoApprox}. In the finely tuned geometries for which $t_c = 0$, we have one joint for all $t_b$ except for at the instantaneous moment $t_b = 0$ at which there are two. Thus, there is essentially no phase transition of the WDW patch for this case.

We can relate the location of the joint to boundary time. Using the expressions for coordinate time along the null ray in \eqref{nullTime}, we write (assuming $|t_b| > t_c$)
\begin{equation}
|t_b| = P\int_0^{r_m} \frac{e^{\chi(r)/2}}{F(r)} dr.
\end{equation}
This can be differentiated to yield
\begin{equation}
\frac{dr_m}{dt_b} = \begin{cases}
e^{-\chi(r_m)/2}F(r_m)&\text{if}\ t_b > t_c,\\
0&\text{if}\ |t_b| < t_c,\\
-e^{-\chi(r_m)/2}F(r_m)&\text{if}\ t_b < -t_c.
\end{cases}\label{drmdtb}
\end{equation}
The emergence and evolution of the joint control the time dependence of complexity; $d\mathcal{C}_A/dt_b \neq 0$ if and only if $|t_b| > t_c$. This is seen through the explicit evaluation of \eqref{complexityWDW} (with appropriate regulator cutoffs) in terms of metric functions. The calculations are performed in AdS-Schwarzschild by \cite{Carmi:2017jqz} but should generalize to the Schwarzschild-like ansatz \eqref{schwarzLike} due to both spherical symmetry and time-reversal symmetry being maintained. The calculation is performed by breaking the WDW patch and its boundary into sections:
\begin{itemize}
\item[(i)] \textbf{$\mathcal{I}_W$ (Bulk):} Consider (I) the future interior piece, (II) the exterior piece, and (III) the past interior piece. (II) is always time-independent, while (I) and (III) are each time-dependent \cite{Carmi:2017jqz}. These time dependencies cancel exactly when the WDW patch reaches both the future and past singularities, which happens if $|t_b| < t_c$ but not if $|t_b| > t_c$.

\item[(ii)] \textbf{$\mathcal{I}_B$ (Boundary):} An appropriate affine parameterization of the normal vectors for the null sheets will make their contributions to the action vanish, so the only boundary terms are those of the regulator surfaces. For $|t_b| < t_c$, we have surfaces near the UV boundary, the future singularity, and the past singularity. Only the latter two contribute time dependence to complexity; their respective time dependencies cancel exactly. When $|t_b| > t_c$, one of the near-singularity surfaces is lost and we are left with the time dependence of the remaining surface.

\item[(iii)] \textbf{$\mathcal{I}_J$ (Joint):} In the regulated calculation, there are joints located both near the UV boundary and at the singularity. The former are time-independent while the latter vanish \cite{Chapman:2016hwi}. The only nontrivial time dependence comes from the joint which probes the trans-IR when $|t_b| > t_c$.
\end{itemize}
We leave the concrete calculation of complexity and its growth rate $d\mathcal{C}_A/dt_B$ to future work,\footnote{Since the first version of this manuscript was written, \cite{Auzzi:2022bfd} has performed this calculation.} but as a preliminary step we compute the action of the joint in Appendix \ref{app:D} and see that it is modified by terms dependent on $a_T$ at the joint. Still, our statement that CA is time-dependent only when probing the trans-IR through its joint relies purely on general symmetry considerations.

\subsubsection{Black Holes without Complementarity}\label{sec:noComp}

The discussion thus far has been about static black holes. However, we can ask about other types of black holes, such as those with angular momentum or charge \cite{Carmi:2017jqz,Auzzi:2018pbc,Bernamonti:2021jyu,Auzzi:2022bfd}. Specifically, we will focus on such cases with hair going to the boundary, so that the geometries may be interpreted as RG flows. \cite{Hartnoll:2020rwq,Cai:2020wrp,Cai:2021obq} indicate that such cases involve black holes without inner horizons, so we assume that the black holes of interest have only one horizon.

Generically, we can have three different types of black hole ``shapes." Each is characterized by the sign of $t_c$, which may be calculated for spherically symmetric black holes using the integral in \eqref{critTimeGeo}:\footnote{The names come from how the future singularities ``bend" with respect to the conformal boundaries \cite{Auzzi:2022bfd}. ``Type D" refers to a future singularity bending down, while ``type U" means that the singularity bends up. ``squarelike" comes from \cite{Fidkowski:2003nf}.}
\begin{align}
t_c > 0 &\implies \text{type D},\\
t_c = 0 &\implies \text{squarelike},\\
t_c < 0 &\implies \text{type U}.
\end{align}
These conditions, respectively, correspond to whether two null geodesics fired from opposite boundaries at the initial slice intersect with the singularity, with one another \textit{at} the singularity, or with each other without reaching the singularity.

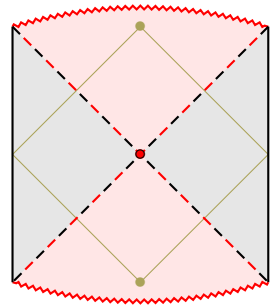
\begin{figure}
\centering
\begin{tikzpicture}[scale=1.7]
\draw[-,draw=none,fill=black!10] (1,1) to (1,-1) to (0,0) to (1,1); 
\draw[-,draw=none,fill=black!10] (-1,1) to (-1,-1) to (0,0) to (-1,1);

\draw[-,draw=none,fill=red!10] (-1,1) ..controls (-0.3,1.2) and (0.3,1.2) .. (1,1) to (0,0) to (-1,1); 
\draw[-,draw=none,fill=red!10] (1,-1) ..controls (0.3,-1.2) and (-0.3,-1.2) .. (-1,-1) to (0,0) to (1,-1);

\draw[-,black!40!yellow] (1,0) to (0,1) to (-1,0) to (0,-1) to (1,0);
\node[black!40!yellow] at (0,1) {$\bullet$};
\node[black!40!yellow] at (0,-1) {$\bullet$};

\draw[-,thick,red,decoration = {zigzag,segment length = 1mm, amplitude = 0.25mm},decorate] (-1,1) ..controls (-0.3,1.2) and (0.3,1.2) .. (1,1);
\draw[-,thick] (1,1) to (1,-1);
\draw[-,thick,red,decoration = {zigzag,segment length = 1mm, amplitude = 0.25mm},decorate] (1,-1) ..controls (0.3,-1.2) and (-0.3,-1.2) .. (-1,-1);
\draw[-,thick] (-1,-1) to (-1,1);

\draw[-,dashed,thick,dash pattern= on 4pt off 8pt,dash phase=6pt,red] (-1,1) to (1,-1);
\draw[-,dashed,thick,dash pattern= on 4pt off 8pt,dash phase=6pt,red] (1,1) to (-1,-1);

\draw[-,dashed,thick,dash pattern= on 4pt off 8pt] (-1,1) to (1,-1);
\draw[-,dashed,thick,dash pattern= on 4pt off 8pt] (1,1) to (-1,-1);

\node[red] at (0,0) {$\bullet$};
\node at (0,0) {$\circ$};
\end{tikzpicture}
\caption{A schematic representation of a type-U black hole. Unlike in type-D or squarelike black holes, null geodesics fired from the boundaries at $t_b = 0$ intersect finitely deep inside of the black hole interior.}
\label{figs:typeU}
\end{figure}

We have been focusing on static black holes with $t_c \geq 0$, i.e. those which are either type D or squarelike. However, it is natural to wonder if type-U solutions (Figure \ref{figs:typeU}) are also possible in that framework. Indeed, type-U solutions are known to exist for charged black holes \cite{Auzzi:2022bfd}.

However, these type-U solutions in practice come from particular deformations of the inner Cauchy horizon.\footnote{We thank Roberto Auzzi for providing us with this point.} Thus, one could in principle get them for the charged and rotating cases. But, since there is no Cauchy horizon at all in the static case, we would expect to only have type-D and squarelike black holes in that case. This is supported by our numerical calculation of the critical time in Appendix \ref{app:C}, for which we only have $t_c > 0$ for a class of free Kasner flows. It would be interesting to further probe this claim.

Observe that complementarity is no longer a feature of the type-U black holes---there is always at least one joint present, and so CA is a good probe of the trans-IR for all boundary times $t_b$. However, there is a continuous window $|t_b| < |t_c|$ in which there are two joints present, with a discontinuous change in the number of joints coinciding with the geodesics becoming nearly null. We leave further exploration of this phenomenon in type-U black holes to future work.

\subsection{Loss of Degrees of Freedom}

A natural question to ask is how quantum information encodes the vanishing of degrees of freedom at the singularity. How this happens is currently unclear. Nonetheless, we provide some preliminary direction towards addressing this question.

First, note that only some probes would actually see $a_T \to 0$---the quantities which actually reach the near-singularity region. This would rule out things like entanglement entropy or CV as encoding the vanishing. However, we would expect other quantities like 2-point correlators or CA to somehow be informed of this phenomenon.

While the details on the quantum information side are vague, the argument on the gravitational side is on more solid footing. Because $a_T(r) \sim e^{-(d-1)\chi(r)/2}$ \eqref{aFunctionSch}, $a_T \to 0$ if $\chi(r)$ diverges in $r$ at least logarithmically. Thus, the matter must source a particularly destructive type of backreaction in the near-singularity regime. Understanding the vanishing through the lens of quantum information goes hand-in-hand with understanding how quantum information encodes near-singularity backreaction of the black hole geometry.

\section{Conclusions}

We have explored trans-IR flows as a framework with which to conceptualize the physics of black hole interiors. We have extended the usual holographic $a$-function to include such analytic continuations of RG flows, proving monotonicity for reasonable deformations (i.e. those dual to matter satisfying the null energy condition) and arguing that the degrees of freedom thin out as we approach the endpoint of the trans-IR---the black hole singularity. We have also embedded the story of quantum information within the trans-IR framework, based on the observation that boundary probes of quantum information are informed by bulk geometry behind the horizon. More specifically, the boundary theory encodes the trans-IR part of the flow as quantum information in various nontrivial ways.

We hope that viewing black hole interiors as trans-IR flows motivates further discovery and exploration of other probes of physics inside of black holes. While our $a_T$-function is one such quantity, one could also consider other measures counting the degrees of freedom, such as the chiral anomaly coefficient in $d = 2$ \cite{Belin:2015jpa} or path integral complexity \cite{Caputa:2017urj,Caputa:2017yrh,Caputa:2021pad}. Conversely, one may explore how other quantum information probes such as R\'enyi entropy, entanglement negativity \cite{Banerjee:2015coc}, and generalized ``complexities" \cite{Belin:2021bga} are informed by our $a_T$-function or the trans-IR regime in general. A parallel illuminating direction would be to study how quantum information encodes the trans-IR using only boundary-theory techniques and calculations, thus moving away from holographic boundary CFTs.

Another quantity often associated with an RG flow is the $\beta$-function of the running coupling. Upon specifying the deformation, one may write $\beta$-functions holographically. This was done for scalar deformations using the superpotential formalism \cite{Kiritsis:2016kog,Gursoy:2018umf}. One could use $\beta$-functions to characterize the dynamics of trans-IR flows.

A natural extension of our work is to consider more types of flows, such as those sourced by combinations of interacting scalars, higher-rank fields, and axions \cite{Wang:2020nkd,Hartnoll:2020fhc,Sword:2021pfm,Mansoori:2021wxf,Das:2021vjf,Liu:2021hap}. One may also consider higher-curvature gravity \cite{Grandi:2021ajl}, but note that we would expect corrections to the $a_T$-function by analogy to the $a$-function of \cite{Myers:2010xs,Myers:2010tj}.

We have only taken a a few steps towards addressing our overall question of what it means to subject a field theory to a trans-IR flow. Nonetheless we envision that understanding this question is equivalent to better understanding physics inside of black holes. We also expect the trans-IR technology to have more applications to the study of more general RG flows involving nonholographic quantum field theories.

\acknowledgments

We thank Sean Hartnoll for early-stage discussion and the name ``trans-IR" and Alexandre Belin for discussion about RG flows towards singularities. We also thank Roberto Auzzi for comments on complexity and the critical time in rotating and charged hairy black holes.
The work of EC and SS is supported by National Science Foundation (NSF) Grant No.
PHY-2112725. AK acknowledges support from the Department of Atomic Energy, Govt. of India, and CEFIPRA grant no. 6304-3. AKP is supported by the Council of Scientific \& Industrial Research (CSIR) Fellowship No. 09/489(0108)/2017-EMR-I.

\begin{appendix}

\section{Characterizing Analytic Continuation of Time to the Interior}\label{app:A}

Our goal in this Appendix is to classify the ambiguity of coordinate time in the interior of asymptotically AdS black holes \eqref{schwarzLike}. This ambiguity depends upon how we choose to analytically continue from exterior time. The interior can thus be thought of as a particular branch of an infinite number of replica geometries, each labeled by a half-integer $\gamma$ \eqref{analyticCont}.

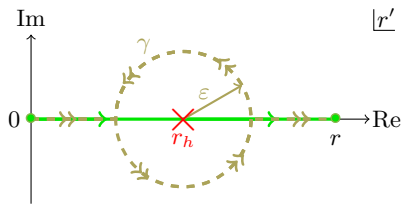
\begin{figure}[t]
\centering
\begin{tikzpicture}[scale=2.25]
\node at (2+0.1,0.6) {$r'$};
\draw[-] (1.925+0.1,0.65) to (1.925+0.1,0.525) to (2.05+0.1,0.525);

\draw[->] (0,0) to (2,0);
\draw[->] (0,-0.5) to (0,0.5);

\draw[-,black!10!green,very thick] (0,0) to (1.8,0);
\draw[->,black!10!green,very thick] (0,0) to (0.45,0);
\draw[->,black!10!green,very thick] (0.9,0) to (1.45,0);

\draw[->>,dashed,black!40!yellow,very thick] (0,0) to (0.27,0);
\draw[-<<,dashed,black!40!yellow,very thick] (0,0) to (0.5,0) arc (180:135:0.4);
\draw[-<<,dashed,black!40!yellow,very thick] (0,0) to (0.5,0) arc (180:45:0.4);
\draw[->>,dashed,black!40!yellow,very thick] (0,0) to (0.5,0) arc (180:0:0.4) to (1.65,0);
\draw[-,dashed,black!40!yellow,very thick] (0,0) to (0.5,0) arc (180:0:0.4) to (1.8,0);

\draw[-<<,dashed,black!40!yellow,very thick] (1.3,0) arc (0:-45:0.4);
\draw[-<<,dashed,black!40!yellow,very thick] (1.3,0) arc (0:-135:0.4);
\draw[-,dashed,black!40!yellow,very thick] (1.3,0) arc (0:-180:0.4);

\node[black!10!green] at (0,0) {$\bullet$};
\node[black!10!green] at (1.8,0) {$\bullet$};

\node[black!40!yellow] at (0,0) {$\circ$};
\node[black!40!yellow] at (1.8,0) {$\circ$};

\draw[->,thick,black!40!yellow] (0.9,0) to (0.9+0.4*0.866,0.4/2);
\node[black!40!yellow] at (0.9+0.4*0.433-0.05,0.2/2+0.05) {$\varepsilon$};

\node[red] at (0.9,0) {\LARGE$\times$};
\node[red] at (0.9,-0.125) {$r_h$};

\node at (0,0.6) {Im};
\node at (2.1,0) {Re};

\node at (-0.1,0) {$0$};
\node at (1.8,-0.125) {$r$};

\node[black!40!yellow] at (0.9-0.4/1.414+0.05,0.4/1.414+0.15) {$\gamma$};
\end{tikzpicture}
\caption{Integration contours used to obtain the interior time coordinate of the null rays \eqref{nullTraj}. The formal, ``bad" integral is taken over the green contour $[0,r]$, but this hits the pole at the red $\times$. Thus, we integrate over a modified contour (in yellow) with half-integer winding number $\gamma$ and take $\varepsilon \to 0$ to separate out the pole, leaving us with the principal part plus an imaginary, residue-dependent, $\gamma$-dependent term \eqref{intPole}.}
\label{figs:contourIntegral}
\end{figure}

We proceed by tracking coordinate time along null rays in \eqref{schwarzLike}. Much of our analysis resembles that of \cite{Fidkowski:2003nf}. The null rays of interest are confined to a constant-$\vec{x}$ slice and are parametrized as $t = t(r)$, where $t(0) = t_b$ is the boundary time. From \eqref{schwarzLike}, the trajectories are
\begin{equation}
\frac{dt}{dr} = \pm \frac{e^{\chi(r)/2}}{F(r)}.\label{nullTraj}
\end{equation}
$+$ denotes the future-directed (``infalling") direction while $-$ denotes the past-directed (``outgoing") direction. Integrating \eqref{nullTraj} yields the future-directed and past-directed trajectories, respectively written as $t_+(r)$ and $t_-(r)$,
\begin{equation}
t_{\pm}(r) = t_b \pm \int_0^{r} \frac{e^{\chi(r')/2}}{F(r')} dr'.\label{nullTime}
\end{equation}
This is well defined up to the horizon $r = r_h$. To reach the interior $r > r_h$, the integral must include the horizon. However, the integrand has a pole here with residue
\begin{equation}
\text{Res}\left[\frac{e^{\chi(r')/2}}{F(r')};r_h\right] = \frac{e^{\chi(r_h)/2}}{F'(r_h)} = -\frac{1}{4\pi T},
\end{equation}
so as in \cite{Hartman:2013qma}, we modify the contour in the complexified $r'$ space (Figure \ref{figs:contourIntegral}) to write
\begin{equation}
\int_0^{r} \frac{e^{\chi(r')/2}}{F(r')} dr' = P\int_0^{r} \frac{e^{\chi(r')/2}}{F(r')} dr' - \frac{\gamma i}{2 T}.\label{intPole}
\end{equation}
The first term is the principal value while the second term is obtained from the pole. Notably, the latter depends on ``how many times" the modified contour goes around the pole to avoid it, captured by the ``winding number" $\gamma$.

The residue theorem is typically used for integration over closed contours, whereby winding numbers are positive integers for counterclockwise closed contours and negative integers for clockwise closed contours. However, we use fractional winding numbers to characterize circular contours which avoid a pole but do not form closed loops. Specifically, a half-integer winding number means that the contour fully wraps around the pole an integer number of times, then travels an additional $\pi$ radians. It is these contours in the complexified $r'$-space which allow us to reach the interior from the exterior.

So, we plug \eqref{intPole} into \eqref{nullTime} to write the coordinate time along a null ray starting at boundary time $t_b$,
\begin{equation}
t_{\pm}(r) = \begin{cases}
t_b \pm {\displaystyle\int_0^r} \dfrac{e^{\chi(r')/2}}{F(r')} dr' &\text{if}\ r < r_h,\vspace{0.2cm}\\
t_b \pm P{\displaystyle\int_0^r} \dfrac{e^{\chi(r')/2}}{F(r')} dr' \mp \dfrac{\gamma i}{2T} &\text{if}\ r > r_h.
\end{cases}\label{trajNull2}
\end{equation}
Coordinate time is real in the exterior, but in the interior time picks up a purely imaginary piece from the pole at the horizon. While we extract this imaginary term from an analysis of null rays, it is also seen in analyses of extremal spacelike surfaces reaching the interior \cite{Hartman:2013qma,Frenkel:2020ysx}. 

Note that $\gamma$ is purely dependent upon how we choose to deform the contour and is thus an inherent ambiguity in time behind the horizon. There are countably infinitely many sheets which we may take to be ``the" interior. Any choice of $\gamma$ is perfectly valid in bulk analyses; earlier work \cite{Fidkowski:2003nf,Hartman:2013qma,Frenkel:2020ysx} just chooses a particular branch $|\gamma| = 1/2$. At least some physics on the boundary should not care about this choice, but additional symmetry is required when computing such physical quantities. For example the critical time $t_c$ (Section \ref{sec:complementarity}) is a $\gamma$-independent quantity found by forcing left-moving and right-moving null rays to meet at $\text{Re}(t) = 0$. One may wonder how the infinitude of choice in $\gamma$ may correspond to infinite families of observables in the boundary theory \cite{Leutheusser:2021qhd}.

We make one more point about the two-sidedness of coordinate time. What we have considered above and throughout this paper is not actually the most generic way to embed two-sided black holes in complex coordinate space. We may instead assume that time is real only in one exterior region. Then, in order to extend time to the other exterior, we must dodge two poles, giving us the freedom of two half-integer winding numbers instead of one. Thus, we may take the other exterior region to still have complex time \cite{Fidkowski:2003nf}. However, assuming symmetric boundary times $t_b$ on both sides of the black hole forces us to keep time real in both exterior regions, which constrains us to just one $\gamma$ for the entire interior.

\section{Review of Extremal Symmetric Spacelike Surfaces}\label{app:B}

In this Appendix, we briefly review extremal symmetric spacelike surfaces of codimension $(k+1)$ in $(d+1)$-dimensional Schwarzschild-like black holes \eqref{schwarzLike}. For $k < d-1$, these surfaces are the sort depicted in Figure \ref{figs:extremalSpacelikeSurfaces} and relevant to the discussion in Section \ref{sec:spacelikeSurfBad}.\footnote{This statement comes with the caveat that there are enough dimensions in the first place for these surfaces to be at least $2$-dimensional, i.e. that they are not geodesics.} Specifically, both the bulk volume slices used in CV ($k = 0$) and HM surfaces ($k = 1$) are of this type. For $k = d-1$, these surfaces are simply the geodesics depicted in Figure \ref{figs:spacelikeGeodesics} and discussed in Section \ref{sec:complementarity}. \cite{Frenkel:2020ysx} also reviews $(k+1)$-codimensional surfaces in \eqref{schwarzLike} with a specific focus on $k = 1$ and $k = d-1$ in $d = 3$.

We parameterize coordinate time on the surface as $t = t(r)$, with boundary time set to $t(0) = t_b$ on both sides of the black hole. Furthermore, we assume $k$ of the components of $\vec{x}$ are constant along the surface. From \eqref{schwarzLike}, the induced metric for $k < d-1$ is then
\begin{equation}
ds_{k}^2 = \frac{1}{r^2}\left[\left(\frac{1}{F(r)} - \frac{F(r)t'(r)^2}{e^{\chi(r)}}\right)dr^2+ \sum_{i=k+1}^{d-1} (dx^i)^2\right],
\end{equation}
where $(x^{k+1},...,x^{d-1})$ are the remaining transverse directions. For $k = d-1$, as all components of $\vec{x}$ are fixed,
\begin{equation}
ds_{d-1}^2 = \frac{1}{r^2}\left(\frac{1}{F(r)} - \frac{F(r)t'(r)^2}{e^{\chi(r)}}\right)dr^2.
\end{equation}
Thus, the coefficient of the volume form for the surface (for both cases) is
\begin{equation}
\mathcal{L}_k = \frac{1}{r^{d-k}}\sqrt{\frac{1}{F(r)} - \frac{F(r)t'(r)^2}{e^{\chi(r)}}}.
\end{equation}
As this surface is anchored to both boundaries, it must achieve some maximal radius $r_m \geq r_h$. By using this and the symmetry, we write the total volume functional as\footnote{This is UV divergent because we are integrating from $r = 0$. We only need to regulate if \eqref{volTot} needs to be explicitly computed.}
\begin{equation}
V_k[t(r)] = 2v_{d-k-1}\int_0^{r_m} \frac{dr}{r^{d-k}}\sqrt{\frac{1}{F(r)} - \frac{F(r)t'(r)^2}{e^{\chi(r)}}},\label{volTot}
\end{equation}
where $v_{d-k-1}$ is the volume of the remaining transverse space ($v_{d-k-1} = \int \prod_{i=k+1}^{d-1} dx^i$ for $k < d-1$ and $v_0 = 1$). Our goal now is to extremize \eqref{volTot}. Because of time-translation symmetry, there is a constant ``energy" $\mathcal{E}$ which is the partial derivative of the integrand with respect to $t'(r)$, so we have\footnote{We have absorbed the sign ambiguity into $\mathcal{E}$.}
\begin{equation}
t'(r) = \frac{\text{sgn}(\mathcal{E}) e^{\chi(r)/2}}{F(r)\sqrt{1 + F(r)e^{-\chi(r)}/(r^{d-k} \mathcal{E})^2}}.\label{derivSp}
\end{equation}
For this extremal surface, while there are three parameters at face value---$t_b$, $r_m$, and $\mathcal{E}$---there are two constraints on them, so there is only one free parameter. The first constraint comes from $r_m$ being the ``turnaround" point of the surface,\footnote{Recall that as $r_m \geq r_h$, $-F(r_m) = |F(r_m)|$.}
\begin{equation}
\frac{1}{t'(r_m)} = 0 \implies \mathcal{E}^2 = \frac{|F(r_m)|e^{-\chi(r_m)}}{r_m^{2(d-k)}}.\label{constraint1}
\end{equation}
As for the second, we integrate \eqref{derivSp} over $r \in [0,r_m]$ to write the formal integral
\begin{equation}
t(r_m) - t_b = \int_0^{r_m} \frac{\text{sgn}(\mathcal{E}) e^{\chi(r)/2}}{F(r)\sqrt{1 + F(r)e^{-\chi(r)}/(r^{d-k} \mathcal{E})^2}} dr.
\end{equation}
This integral is divergent; its integrand has a simple pole at $r = r_h$. As discussed in detail in Appendix \ref{app:A}, we slightly deform the contour $r \in [0,r_m]$ into complexified $r$-space to avoid this pole, picking up an ambiguous imaginary term in the process \eqref{analyticCont}. Fortunately, the symmetry of these particular surfaces implies $\text{Re}[t(r_m)] = 0$, so this ambiguity does not matter and we are left with the principal value,
\begin{equation}
t_b = -P\int_0^{r_m} \frac{\text{sgn}(\mathcal{E}) e^{\chi(r)/2}}{F(r)\sqrt{1 + F(r)e^{-\chi(r)}/(r^{d-k} \mathcal{E})^2}} dr.\label{constraint2}
\end{equation}
\eqref{constraint1} and \eqref{constraint2} are also in \cite{Frenkel:2020ysx}. The main point is that fixing any one of $t_b$, $r_m$, and $\mathcal{E}$ will fix the rest. Notably,
\begin{equation}
t_b = 0 \iff \mathcal{E} = 0 \iff r_m = r_h.
\end{equation}
Now, plugging \eqref{derivSp} into \eqref{volTot}, we get that the volume ``density" $\mathcal{V}_k = V_k/v_{d-k-1}$ of the surface of energy $\mathcal{E}$ is
\begin{equation}
\mathcal{V}_k(\mathcal{E}) = \frac{2}{|\mathcal{E}|}\int_0^{r_m} \frac{dr}{r^{2(d-k)}} \frac{e^{-\chi(r)/2}}{\sqrt{1 + F(r)e^{-\chi(r)}/(r^{d-k}\mathcal{E})^2}}.\label{volk}
\end{equation}

\subsection{Surface Barriers in Kasner Flows}\label{app:B1}

Our claim that entanglement entropy and CV generally fail to probe the full trans-IR regime (Section \ref{sec:spacelikeSurfBad}) rest on the appropriate $k < d-1$ surfaces being blocked by an extremal surface barrier. While \cite{Engelhardt:2013tra} claims that such barriers are generic features of asymptotically locally AdS black holes, demonstrating existence is more straightforward upon specifying the geometry \cite{Hartman:2013qma,Frenkel:2020ysx}. We find it instructive to see how such barriers are found in Kasner flows, following \cite{Frenkel:2020ysx}.

We start with the function
\begin{equation}
g(r) = \frac{F(r)e^{-\chi(r)}}{r^{2(d-k)}}.
\end{equation}
Clearly $g(r_h) = 0$. Furthermore, as $r \to \infty$, we plug in the asymptotic behavior of the metric functions \eqref{asymptChi} and \eqref{asymptF} to write
\begin{equation}
g(r) \sim -f_1 e^{-\chi_1}r^{q'},\ \ r \to \infty,
\end{equation}
where we have defined
\begin{equation}
q' = 2k-d-\frac{1}{2}(d-1)q^2.
\end{equation}
As $g(r)$ follows a power rule, it is straightforward to deduce its limit as $r \to \infty$.
\begin{equation}
\lim_{r \to \infty} g(r) = \begin{cases}
-\infty&\text{if}\ q' > 0,\\
-f_1 e^{-\chi_1}&\text{if}\ q' = 0,\\
0&\text{if}\ q' < 0.
\end{cases}
\end{equation}
However, so long as $k < d-1$ (i.e. so long as the surface is not just a geodesic),
\begin{equation}
q' < k - 1 - \frac{1}{2}(d-1)q^2,
\end{equation}
from which we deduce that
\begin{equation}
k = 0,1 \implies q' < 0.
\end{equation}
Thus, in general for the surfaces relevant to entanglement entropy and CV, $g(r)$ is $0$ both at the horizon and at the singularity. Furthermore, it is clearly negative in the interior, so there must be a finite $r_c \in (r_h,\infty)$ such that
\begin{equation}
g(r_c) = \min_{r \in (r_h,\infty)} g(r).
\end{equation}
From \eqref{constraint1}, we have a ``critical" energy $\mathcal{E}_c$ for which $g(r_c) = -\mathcal{E}_c^2$, so
\begin{equation}
1 + \frac{g(r_c)}{\mathcal{E}_c^2} = 1 + \frac{F(r_c)e^{-\chi(r_c)}}{(r_c^{d-k}\mathcal{E}_c)^2} = 0.
\end{equation}
Now consider the boundary time $t_b$ in \eqref{constraint2} as we take $r_m \to r_c$ and $\mathcal{E} \to \mathcal{E}_c$. The denominator of the integrand goes to $0$, so the integral itself diverges. In other words, as we take $t_b \to \infty$, the corresponding $k = 0,1$ surfaces will get ``stuck" at $r_m = r_c < \infty$.

In this analysis, one may ask about larger $k > 1$ which still satisfy $k < d-1$. Applying the arguments above becomes muddier because we may get $q' > 0$ for perfectly valid geometries, in spite of the generic statements of \cite{Engelhardt:2013tra}. For example, consider a Schwarzschild singularity, which we recover from Kasner by setting $q^2 = 0$. Then,
\begin{equation}
q' = 2k - d.
\end{equation}
This is positive precisely when $k > d/2$, and there are integer choices of $k$ and $d$ which fit into this range, such as $k = 3$ and $d = 5$ which would yield $q' = 1$. Thus we cannot assert the existence of a finite $r_c$ at which $g(r)$ is minimized.

The argument above fails to detect an extremal surface barrier for particular lower-dimensional surfaces in certain geometries. Nonetheless, this does not matter for entanglement entropy ($k = 1$) and CV ($k = 0$), so the main claim of Section \ref{sec:spacelikeSurfBad} is undamaged.

\subsection{Late-Time Length and ${\boldsymbol{a_T}}$}\label{app:B2}

For geodesic lengths $k = d-1$, the maximal radius $r_m$ reaches all the way to the singularity as $t_b$ approaches the critical time $t_c$ discussed in Section \ref{sec:Q3}. In the main text, we focus on $r_m$ as a basic measure of how far the symmetric 2-point correlator dual to a geodesic probes into the trans-IR. Here we go further with this discussion, explaining how near-critical-$t_b$ (that is, $t_b \sim t_c$) geodesic lengths $L$ are informed by the $a_T$-function \eqref{aFunctionSch} and deeper geometrical data describing the trans-IR flow.

The formal length (\eqref{volk} with $k = d-1$) at near-critical times is found from a large-$|\mathcal{E}|$ expansion,
\begin{equation}
L \sim \frac{2}{|\mathcal{E}|} \int_0^{r_m} \frac{e^{-\chi(r)/2}}{r^2}dr,\ \ t_b \to t_c.
\end{equation}
Up to a factor of $r^2$, the integrand is simply a power of the $a_T$-function \eqref{aFunctionSch}---omitting the constant prefactors,
\begin{equation}
L \sim \frac{1}{|\mathcal{E}|} \int_0^{r_m} \frac{a_T(r)^{1/(d-1)}}{r^2} dr.
\end{equation}
There is a geometrical interpretation of the integrand in terms of constant-$t$ and constant-$r$ slices. Their respective unit normal 1-forms may be written down as follows:
\begin{align}
\widetilde{\alpha}_t = \frac{\sqrt{F(r)}}{r}e^{-\chi(r)/2} dt,\ \ \widetilde{\alpha}_r = \frac{1}{r\sqrt{F(r)}} dr.
\end{align}
We then take the wedge product, producing a 2-form,
\begin{equation}
\widetilde{\alpha}_t \wedge \widetilde{\alpha}_r = \frac{e^{-\chi(r)/2}}{r^2} dt \wedge dr.
\end{equation}
This is precisely the volume form of the $(t,r)$ plane. Thus $L$ in the regime $t_b \to t_c$ is, up to an overall prefactor, the volume ``density" with respect to $t$ of the $(t,r)$ plane (i.e. the volume divided by the size of $t$ space $\int_{\mathbb{R}} dt$).

\section{The Critical Time in Free Kasner Flows}\label{app:C}

The discussion of Section \ref{sec:complementarity} regarding the complementarity of the 2-point correlator and CA is rather general. Here, we focus on free Kasner flows, whose dynamics are controlled by \eqref{eom1}--\eqref{eom3}. Taking $d = 3$ and $\Delta = 2$ for concreteness, we plot the critical time $t_c$ as a function of the deformation parameter $\phi_0/T$.

Following the procedure of shooting null rays from opposite boundaries at the same boundary time $t_b$, we have that the critical time $t_c$ is
\begin{equation}
t_c = P\int_0^{\infty} \frac{e^{\chi(r)/2}}{F(r)}dr.
\end{equation}
Generically, as $t_c$ is an integral of metric functions, it cannot be calculated in closed form for arbitrary geometries. However, we can still do so for planar AdS-Schwarzschild black holes, in which $\chi(r) = 0$ and $F(r) = 1-(r/r_h)^d$. Careful consideration is needed to evaluate the principal value, but we find that
\begin{equation}
t_c = P\int_0^\infty \frac{dr}{1-(r/r_h)^d} \implies \frac{t_c}{r_h} = \frac{\pi}{d}\cot\left(\frac{\pi}{d}\right).\label{adsSchwarzCrit}
\end{equation}
This coincides with the value in \cite{Carmi:2017jqz} upon appropriate synchronization of conventions; their conventions produce an extra factor of $2$.

\begin{figure}
\centering
\includegraphics[scale=0.65]{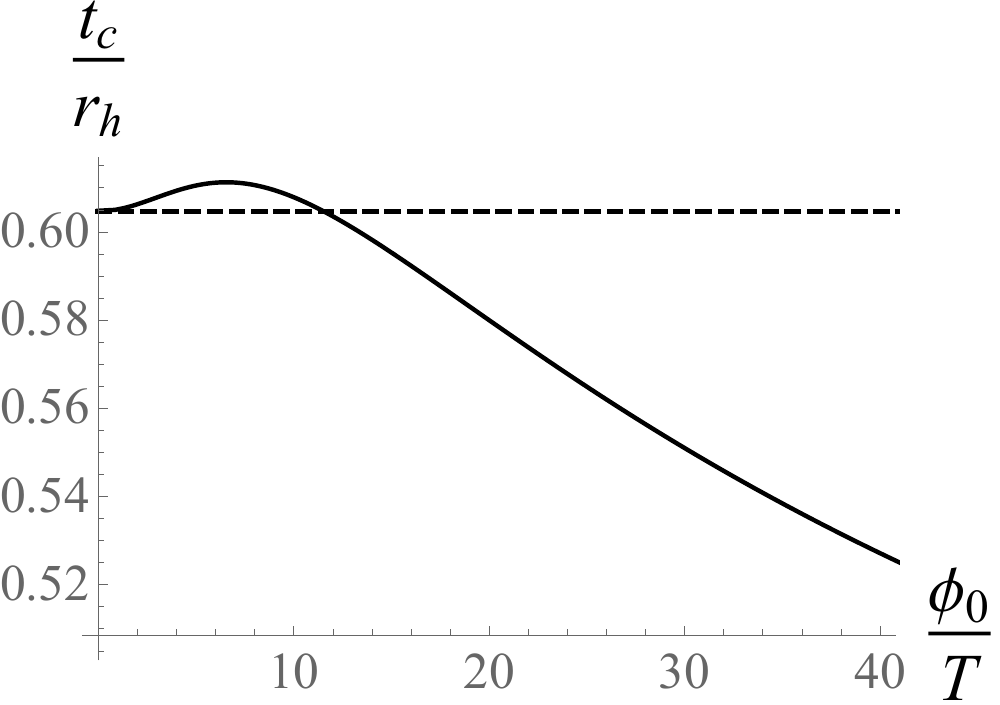}
\caption{The critical time $t_c$ (in units of $r_h$) characterizing the complementarity of the 2-point correlator and CA, plotted as a function of the deformation parameter $\phi_0/T$ for free Kasner flows with $d = 3$, $\Delta = 2$. The value at $\phi_0/T = 0$ coincides with that of AdS-Schwarzschild, $t_c/r_h = \pi/(3\sqrt{3})$. $t_c/r_h$ has a maximum value of $0.61$ at $\phi_0/T \approx 6.6$. As it decreases, $t_c/r_h$ again attains the AdS-Schwarzschild value at $\phi_0/T \approx 11.5$.}
\label{figs:critTimeVsDef}
\end{figure}

Using the numerical Kasner solutions to \eqref{eom1}--\eqref{eom3}, we may still solve for $t_c$ as a function of the deformation parameter $\phi_0/T$ (Figure \ref{figs:critTimeVsDef}).

As expected, $t_c$ starts at the AdS-Schwarzschild value when $\phi_0/T = 0$, but $t_c$ also exhibits nontrivial behavior---rising to a maximum before falling past the AdS-Schwarzschild value again. This is rising and falling is similar to the behavior in the near-singularity Kasner exponent $p_t$ \cite{Frenkel:2020ysx,Caceres:2021fuw}, but our numerics indicate that $t_c$ monotonically decreases forever.

It is unclear if $t_c \to 0$ or asymptotes to a finite value as $\phi_0/T \to \infty$, but either way the highly deformed free Kasner flows in this regime have much more ``squarelike" \cite{Fidkowski:2003nf} Penrose diagrams than AdS-Schwarzschild. This is in spite of the evidence that the endpoint of the trans-IR $\phi_0/T \to \infty$ flow is the Schwarzschild singularity, i.e. that the Kasner exponents approach their Schwarzschild values for large deformations \cite{Frenkel:2020ysx}.

\section{Action at the Joint}\label{app:D}

We compute the contribution of the joint term to the overall complexity---a first step towards obtaining the explicit time dependence.\footnote{\cite{Auzzi:2022bfd} has taken the remainder of these steps since earlier versions of this manuscript were written.} We focus on $t_b > t_c$ (i.e. the null joint in the past interior); the same result holds for $t_b < -t_c$ by symmetry. From \cite{Lehner:2016vdi,Carmi:2017jqz}, the joint term is
\begin{equation}
I_{\mathcal{J}} = \frac{1}{8\pi G_N} \oint \bar{a}\,dS,\ \ \bar{a} = \log\left(-\frac{1}{2}\vec{k}_R \cdot \vec{k}_L\right).
\end{equation}
$\vec{k}_R$ and $\vec{k}_L$ are the outward-directed normal vectors, respectively, for the right past-directed null sheet and the left past-directed null sheet bounding $\mathcal{W}(t_b)$. Analogously to \cite{Carmi:2017jqz} we write their dual 1-forms as
\begin{align}
\widetilde{k}_R &= -\mathcal{N}\left(dt + \frac{e^{\chi(r)/2}}{F(r)} dr\right),\\
\widetilde{k}_L &= \mathcal{N}\left(dt - \frac{e^{\chi(r)/2}}{F(r)} dr\right).
\end{align}
Here, $\mathcal{N} > 0$ is a normalization coefficient which remains ambiguous in the problem. Thus, in the interior,
\begin{equation}
\bar{a} = -\log\left(\frac{|F(r)|}{\mathcal{N}^2 r^2}\right) + \chi(r).
\end{equation}
Integrating $\bar{a}$ on the joint yields a factor of $V_{d-1}/r_m^{d-1}$ where $V_{d-1}$ is the volume of the transverse $\vec{x}$ space, so
\begin{equation}
I_{\mathcal{J}} = -\frac{V_{d-1}}{8\pi G_N r_{m}^{d-1}} \left[\log\left(\frac{|F(r_m)|}{\mathcal{N}^2 r_m^2}\right) - \chi(r_m)\right].
\end{equation}
This agrees with \cite{Carmi:2017jqz}, except there is an additional term corresponding $\chi$ appearing due to backreaction. This is essentially the $a_T$ dependence, since
\begin{equation}
\chi(r_m) = \frac{2}{d-1}\log\left[\frac{\pi^{d/2}}{\Gamma\left(\frac{d}{2}\right)\ell_P^{d-1}}\right] - \frac{2}{d-1}\log a_T(r_m).
\end{equation}
Thus, the time dependence of the joint term depends on $a_T$ and its derivative. Specifically,
\begin{equation}
\frac{d\chi(r_m)}{dt_b} = -\frac{2}{d-1}\frac{dr_m}{dt_b}\left(\frac{1}{a_T} \frac{da_T}{dr_m}\right),
\end{equation}
where $dr_m/dt_b$ is written in \eqref{drmdtb}.
\vfill
\end{appendix}
\pagebreak
\bibliographystyle{apsrev4-2}
\bibliography{multi.bib}

\begin{thebibliography}{76}%
\makeatletter
\providecommand \@ifxundefined [1]{%
 \@ifx{#1\undefined}
}%
\providecommand \@ifnum [1]{%
 \ifnum #1\expandafter \@firstoftwo
 \else \expandafter \@secondoftwo
 \fi
}%
\providecommand \@ifx [1]{%
 \ifx #1\expandafter \@firstoftwo
 \else \expandafter \@secondoftwo
 \fi
}%
\providecommand \natexlab [1]{#1}%
\providecommand \enquote  [1]{``#1''}%
\providecommand \bibnamefont  [1]{#1}%
\providecommand \bibfnamefont [1]{#1}%
\providecommand \citenamefont [1]{#1}%
\providecommand \href@noop [0]{\@secondoftwo}%
\providecommand \href [0]{\begingroup \@sanitize@url \@href}%
\providecommand \@href[1]{\@@startlink{#1}\@@href}%
\providecommand \@@href[1]{\endgroup#1\@@endlink}%
\providecommand \@sanitize@url [0]{\catcode `\\12\catcode `\$12\catcode
  `\&12\catcode `\#12\catcode `\^12\catcode `\_12\catcode `\%12\relax}%
\providecommand \@@startlink[1]{}%
\providecommand \@@endlink[0]{}%
\providecommand \url  [0]{\begingroup\@sanitize@url \@url }%
\providecommand \@url [1]{\endgroup\@href {#1}{\urlprefix }}%
\providecommand \urlprefix  [0]{URL }%
\providecommand \Eprint [0]{\href }%
\providecommand \doibase [0]{https://doi.org/}%
\providecommand \selectlanguage [0]{\@gobble}%
\providecommand \bibinfo  [0]{\@secondoftwo}%
\providecommand \bibfield  [0]{\@secondoftwo}%
\providecommand \translation [1]{[#1]}%
\providecommand \BibitemOpen [0]{}%
\providecommand \bibitemStop [0]{}%
\providecommand \bibitemNoStop [0]{.\EOS\space}%
\providecommand \EOS [0]{\spacefactor3000\relax}%
\providecommand \BibitemShut  [1]{\csname bibitem#1\endcsname}%
\let\auto@bib@innerbib\@empty
\bibitem [{\citenamefont {Frenkel}\ \emph {et~al.}(2020)\citenamefont
  {Frenkel}, \citenamefont {Hartnoll}, \citenamefont {Kruthoff},\ and\
  \citenamefont {Shi}}]{Frenkel:2020ysx}%
  \BibitemOpen
  \bibfield  {author} {\bibinfo {author} {\bibfnamefont {A.}~\bibnamefont
  {Frenkel}}, \bibinfo {author} {\bibfnamefont {S.~A.}\ \bibnamefont
  {Hartnoll}}, \bibinfo {author} {\bibfnamefont {J.}~\bibnamefont {Kruthoff}},\
  and\ \bibinfo {author} {\bibfnamefont {Z.~D.}\ \bibnamefont {Shi}},\ }\href
  {https://doi.org/10.1007/JHEP08(2020)003} {\bibfield  {journal} {\bibinfo
  {journal} {JHEP}\ }\textbf {\bibinfo {volume} {08}},\ \bibinfo {pages}
  {003}},\ \Eprint {https://arxiv.org/abs/2004.01192} {arXiv:2004.01192
  [hep-th]} \BibitemShut {NoStop}%
\bibitem [{\citenamefont {Maldacena}(1998)}]{Maldacena:1997re}%
  \BibitemOpen
  \bibfield  {author} {\bibinfo {author} {\bibfnamefont {J.~M.}\ \bibnamefont
  {Maldacena}},\ }\href {https://doi.org/10.1023/A:1026654312961} {\bibfield
  {journal} {\bibinfo  {journal} {Adv. Theor. Math. Phys.}\ }\textbf {\bibinfo
  {volume} {2}},\ \bibinfo {pages} {231} (\bibinfo {year} {1998})},\ \Eprint
  {https://arxiv.org/abs/hep-th/9711200} {arXiv:hep-th/9711200} \BibitemShut
  {NoStop}%
\bibitem [{\citenamefont {Fidkowski}\ \emph {et~al.}(2004)\citenamefont
  {Fidkowski}, \citenamefont {Hubeny}, \citenamefont {Kleban},\ and\
  \citenamefont {Shenker}}]{Fidkowski:2003nf}%
  \BibitemOpen
  \bibfield  {author} {\bibinfo {author} {\bibfnamefont {L.}~\bibnamefont
  {Fidkowski}}, \bibinfo {author} {\bibfnamefont {V.}~\bibnamefont {Hubeny}},
  \bibinfo {author} {\bibfnamefont {M.}~\bibnamefont {Kleban}},\ and\ \bibinfo
  {author} {\bibfnamefont {S.}~\bibnamefont {Shenker}},\ }\href
  {https://doi.org/10.1088/1126-6708/2004/02/014} {\bibfield  {journal}
  {\bibinfo  {journal} {JHEP}\ }\textbf {\bibinfo {volume} {02}},\ \bibinfo
  {pages} {014}},\ \Eprint {https://arxiv.org/abs/hep-th/0306170}
  {arXiv:hep-th/0306170} \BibitemShut {NoStop}%
\bibitem [{\citenamefont {Festuccia}\ and\ \citenamefont
  {Liu}(2006)}]{Festuccia:2005pi}%
  \BibitemOpen
  \bibfield  {author} {\bibinfo {author} {\bibfnamefont {G.}~\bibnamefont
  {Festuccia}}\ and\ \bibinfo {author} {\bibfnamefont {H.}~\bibnamefont
  {Liu}},\ }\href {https://doi.org/10.1088/1126-6708/2006/04/044} {\bibfield
  {journal} {\bibinfo  {journal} {JHEP}\ }\textbf {\bibinfo {volume} {04}},\
  \bibinfo {pages} {044}},\ \Eprint {https://arxiv.org/abs/hep-th/0506202}
  {arXiv:hep-th/0506202} \BibitemShut {NoStop}%
\bibitem [{\citenamefont {Balasubramanian}\ \emph {et~al.}(2020)\citenamefont
  {Balasubramanian}, \citenamefont {Kar},\ and\ \citenamefont
  {S\'arosi}}]{Balasubramanian:2019qwk}%
  \BibitemOpen
  \bibfield  {author} {\bibinfo {author} {\bibfnamefont {V.}~\bibnamefont
  {Balasubramanian}}, \bibinfo {author} {\bibfnamefont {A.}~\bibnamefont
  {Kar}},\ and\ \bibinfo {author} {\bibfnamefont {G.}~\bibnamefont
  {S\'arosi}},\ }\href {https://doi.org/10.1007/JHEP06(2020)054} {\bibfield
  {journal} {\bibinfo  {journal} {JHEP}\ }\textbf {\bibinfo {volume} {06}},\
  \bibinfo {pages} {054}},\ \Eprint {https://arxiv.org/abs/1911.12413}
  {arXiv:1911.12413 [hep-th]} \BibitemShut {NoStop}%
\bibitem [{\citenamefont {Grinberg}\ and\ \citenamefont
  {Maldacena}(2021)}]{Grinberg:2020fdj}%
  \BibitemOpen
  \bibfield  {author} {\bibinfo {author} {\bibfnamefont {M.}~\bibnamefont
  {Grinberg}}\ and\ \bibinfo {author} {\bibfnamefont {J.}~\bibnamefont
  {Maldacena}},\ }\href {https://doi.org/10.1007/JHEP03(2021)131} {\bibfield
  {journal} {\bibinfo  {journal} {JHEP}\ }\textbf {\bibinfo {volume} {03}},\
  \bibinfo {pages} {131}},\ \Eprint {https://arxiv.org/abs/2011.01004}
  {arXiv:2011.01004 [hep-th]} \BibitemShut {NoStop}%
\bibitem [{\citenamefont {Henningson}\ and\ \citenamefont
  {Skenderis}(1998)}]{Henningson:1998gx}%
  \BibitemOpen
  \bibfield  {author} {\bibinfo {author} {\bibfnamefont {M.}~\bibnamefont
  {Henningson}}\ and\ \bibinfo {author} {\bibfnamefont {K.}~\bibnamefont
  {Skenderis}},\ }\href {https://doi.org/10.1088/1126-6708/1998/07/023}
  {\bibfield  {journal} {\bibinfo  {journal} {JHEP}\ }\textbf {\bibinfo
  {volume} {07}},\ \bibinfo {pages} {023}},\ \Eprint
  {https://arxiv.org/abs/hep-th/9806087} {arXiv:hep-th/9806087} \BibitemShut
  {NoStop}%
\bibitem [{\citenamefont {Freedman}\ \emph {et~al.}(1999)\citenamefont
  {Freedman}, \citenamefont {Gubser}, \citenamefont {Pilch},\ and\
  \citenamefont {Warner}}]{Freedman:1999gp}%
  \BibitemOpen
  \bibfield  {author} {\bibinfo {author} {\bibfnamefont {D.~Z.}\ \bibnamefont
  {Freedman}}, \bibinfo {author} {\bibfnamefont {S.~S.}\ \bibnamefont
  {Gubser}}, \bibinfo {author} {\bibfnamefont {K.}~\bibnamefont {Pilch}},\ and\
  \bibinfo {author} {\bibfnamefont {N.~P.}\ \bibnamefont {Warner}},\ }\href
  {https://doi.org/10.4310/ATMP.1999.v3.n2.a7} {\bibfield  {journal} {\bibinfo
  {journal} {Adv. Theor. Math. Phys.}\ }\textbf {\bibinfo {volume} {3}},\
  \bibinfo {pages} {363} (\bibinfo {year} {1999})},\ \Eprint
  {https://arxiv.org/abs/hep-th/9904017} {arXiv:hep-th/9904017} \BibitemShut
  {NoStop}%
\bibitem [{\citenamefont {Myers}\ and\ \citenamefont
  {Sinha}(2010)}]{Myers:2010xs}%
  \BibitemOpen
  \bibfield  {author} {\bibinfo {author} {\bibfnamefont {R.~C.}\ \bibnamefont
  {Myers}}\ and\ \bibinfo {author} {\bibfnamefont {A.}~\bibnamefont {Sinha}},\
  }\href {https://doi.org/10.1103/PhysRevD.82.046006} {\bibfield  {journal}
  {\bibinfo  {journal} {Phys. Rev. D}\ }\textbf {\bibinfo {volume} {82}},\
  \bibinfo {pages} {046006} (\bibinfo {year} {2010})},\ \Eprint
  {https://arxiv.org/abs/1006.1263} {arXiv:1006.1263 [hep-th]} \BibitemShut
  {NoStop}%
\bibitem [{\citenamefont {Myers}\ and\ \citenamefont
  {Sinha}(2011)}]{Myers:2010tj}%
  \BibitemOpen
  \bibfield  {author} {\bibinfo {author} {\bibfnamefont {R.~C.}\ \bibnamefont
  {Myers}}\ and\ \bibinfo {author} {\bibfnamefont {A.}~\bibnamefont {Sinha}},\
  }\href {https://doi.org/10.1007/JHEP01(2011)125} {\bibfield  {journal}
  {\bibinfo  {journal} {JHEP}\ }\textbf {\bibinfo {volume} {01}},\ \bibinfo
  {pages} {125}},\ \Eprint {https://arxiv.org/abs/1011.5819} {arXiv:1011.5819
  [hep-th]} \BibitemShut {NoStop}%
\bibitem [{\citenamefont {Caceres}\ \emph {et~al.}(2021)\citenamefont
  {Caceres}, \citenamefont {Kundu}, \citenamefont {Patra},\ and\ \citenamefont
  {Shashi}}]{Caceres:2021fuw}%
  \BibitemOpen
  \bibfield  {author} {\bibinfo {author} {\bibfnamefont {E.}~\bibnamefont
  {Caceres}}, \bibinfo {author} {\bibfnamefont {A.}~\bibnamefont {Kundu}},
  \bibinfo {author} {\bibfnamefont {A.~K.}\ \bibnamefont {Patra}},\ and\
  \bibinfo {author} {\bibfnamefont {S.}~\bibnamefont {Shashi}},\ }\href@noop {}
  {\  (\bibinfo {year} {2021})},\ \Eprint {https://arxiv.org/abs/2107.00022}
  {arXiv:2107.00022 [hep-th]} \BibitemShut {NoStop}%
\bibitem [{\citenamefont {Ryu}\ and\ \citenamefont
  {Takayanagi}(2006)}]{Ryu:2006bv}%
  \BibitemOpen
  \bibfield  {author} {\bibinfo {author} {\bibfnamefont {S.}~\bibnamefont
  {Ryu}}\ and\ \bibinfo {author} {\bibfnamefont {T.}~\bibnamefont
  {Takayanagi}},\ }\href {https://doi.org/10.1103/PhysRevLett.96.181602}
  {\bibfield  {journal} {\bibinfo  {journal} {Phys. Rev. Lett.}\ }\textbf
  {\bibinfo {volume} {96}},\ \bibinfo {pages} {181602} (\bibinfo {year}
  {2006})},\ \Eprint {https://arxiv.org/abs/hep-th/0603001}
  {arXiv:hep-th/0603001} \BibitemShut {NoStop}%
\bibitem [{\citenamefont {Hubeny}\ \emph {et~al.}(2007)\citenamefont {Hubeny},
  \citenamefont {Rangamani},\ and\ \citenamefont {Takayanagi}}]{Hubeny:2007xt}%
  \BibitemOpen
  \bibfield  {author} {\bibinfo {author} {\bibfnamefont {V.~E.}\ \bibnamefont
  {Hubeny}}, \bibinfo {author} {\bibfnamefont {M.}~\bibnamefont {Rangamani}},\
  and\ \bibinfo {author} {\bibfnamefont {T.}~\bibnamefont {Takayanagi}},\
  }\href {https://doi.org/10.1088/1126-6708/2007/07/062} {\bibfield  {journal}
  {\bibinfo  {journal} {JHEP}\ }\textbf {\bibinfo {volume} {07}},\ \bibinfo
  {pages} {062}},\ \Eprint {https://arxiv.org/abs/0705.0016} {arXiv:0705.0016
  [hep-th]} \BibitemShut {NoStop}%
\bibitem [{\citenamefont {Susskind}(2016{\natexlab{a}})}]{Susskind:2014rva}%
  \BibitemOpen
  \bibfield  {author} {\bibinfo {author} {\bibfnamefont {L.}~\bibnamefont
  {Susskind}},\ }\href {https://doi.org/10.1002/prop.201500092} {\bibfield
  {journal} {\bibinfo  {journal} {Fortsch. Phys.}\ }\textbf {\bibinfo {volume}
  {64}},\ \bibinfo {pages} {24} (\bibinfo {year} {2016}{\natexlab{a}})},\
  \bibinfo {note} {[Addendum: Fortsch.Phys. 64, 44--48 (2016)]},\ \Eprint
  {https://arxiv.org/abs/1402.5674} {arXiv:1402.5674 [hep-th]} \BibitemShut
  {NoStop}%
\bibitem [{\citenamefont {Stanford}\ and\ \citenamefont
  {Susskind}(2014)}]{Stanford:2014jda}%
  \BibitemOpen
  \bibfield  {author} {\bibinfo {author} {\bibfnamefont {D.}~\bibnamefont
  {Stanford}}\ and\ \bibinfo {author} {\bibfnamefont {L.}~\bibnamefont
  {Susskind}},\ }\href {https://doi.org/10.1103/PhysRevD.90.126007} {\bibfield
  {journal} {\bibinfo  {journal} {Phys. Rev. D}\ }\textbf {\bibinfo {volume}
  {90}},\ \bibinfo {pages} {126007} (\bibinfo {year} {2014})},\ \Eprint
  {https://arxiv.org/abs/1406.2678} {arXiv:1406.2678 [hep-th]} \BibitemShut
  {NoStop}%
\bibitem [{\citenamefont {Susskind}(2016{\natexlab{b}})}]{Susskind:2014moa}%
  \BibitemOpen
  \bibfield  {author} {\bibinfo {author} {\bibfnamefont {L.}~\bibnamefont
  {Susskind}},\ }\href {https://doi.org/10.1002/prop.201500095} {\bibfield
  {journal} {\bibinfo  {journal} {Fortsch. Phys.}\ }\textbf {\bibinfo {volume}
  {64}},\ \bibinfo {pages} {49} (\bibinfo {year} {2016}{\natexlab{b}})},\
  \Eprint {https://arxiv.org/abs/1411.0690} {arXiv:1411.0690 [hep-th]}
  \BibitemShut {NoStop}%
\bibitem [{\citenamefont {Balasubramanian}\ and\ \citenamefont
  {Ross}(2000)}]{Balasubramanian:1999zv}%
  \BibitemOpen
  \bibfield  {author} {\bibinfo {author} {\bibfnamefont {V.}~\bibnamefont
  {Balasubramanian}}\ and\ \bibinfo {author} {\bibfnamefont {S.~F.}\
  \bibnamefont {Ross}},\ }\href {https://doi.org/10.1103/PhysRevD.61.044007}
  {\bibfield  {journal} {\bibinfo  {journal} {Phys. Rev. D}\ }\textbf {\bibinfo
  {volume} {61}},\ \bibinfo {pages} {044007} (\bibinfo {year} {2000})},\
  \Eprint {https://arxiv.org/abs/hep-th/9906226} {arXiv:hep-th/9906226}
  \BibitemShut {NoStop}%
\bibitem [{\citenamefont {Brown}\ \emph
  {et~al.}(2016{\natexlab{a}})\citenamefont {Brown}, \citenamefont {Roberts},
  \citenamefont {Susskind}, \citenamefont {Swingle},\ and\ \citenamefont
  {Zhao}}]{Brown:2015bva}%
  \BibitemOpen
  \bibfield  {author} {\bibinfo {author} {\bibfnamefont {A.~R.}\ \bibnamefont
  {Brown}}, \bibinfo {author} {\bibfnamefont {D.~A.}\ \bibnamefont {Roberts}},
  \bibinfo {author} {\bibfnamefont {L.}~\bibnamefont {Susskind}}, \bibinfo
  {author} {\bibfnamefont {B.}~\bibnamefont {Swingle}},\ and\ \bibinfo {author}
  {\bibfnamefont {Y.}~\bibnamefont {Zhao}},\ }\href
  {https://doi.org/10.1103/PhysRevLett.116.191301} {\bibfield  {journal}
  {\bibinfo  {journal} {Phys. Rev. Lett.}\ }\textbf {\bibinfo {volume} {116}},\
  \bibinfo {pages} {191301} (\bibinfo {year} {2016}{\natexlab{a}})},\ \Eprint
  {https://arxiv.org/abs/1509.07876} {arXiv:1509.07876 [hep-th]} \BibitemShut
  {NoStop}%
\bibitem [{\citenamefont {Brown}\ \emph
  {et~al.}(2016{\natexlab{b}})\citenamefont {Brown}, \citenamefont {Roberts},
  \citenamefont {Susskind}, \citenamefont {Swingle},\ and\ \citenamefont
  {Zhao}}]{Brown:2015lvg}%
  \BibitemOpen
  \bibfield  {author} {\bibinfo {author} {\bibfnamefont {A.~R.}\ \bibnamefont
  {Brown}}, \bibinfo {author} {\bibfnamefont {D.~A.}\ \bibnamefont {Roberts}},
  \bibinfo {author} {\bibfnamefont {L.}~\bibnamefont {Susskind}}, \bibinfo
  {author} {\bibfnamefont {B.}~\bibnamefont {Swingle}},\ and\ \bibinfo {author}
  {\bibfnamefont {Y.}~\bibnamefont {Zhao}},\ }\href
  {https://doi.org/10.1103/PhysRevD.93.086006} {\bibfield  {journal} {\bibinfo
  {journal} {Phys. Rev. D}\ }\textbf {\bibinfo {volume} {93}},\ \bibinfo
  {pages} {086006} (\bibinfo {year} {2016}{\natexlab{b}})},\ \Eprint
  {https://arxiv.org/abs/1512.04993} {arXiv:1512.04993 [hep-th]} \BibitemShut
  {NoStop}%
\bibitem [{\citenamefont {Balasubramanian}\ and\ \citenamefont
  {Kraus}(1999)}]{Balasubramanian:1999jd}%
  \BibitemOpen
  \bibfield  {author} {\bibinfo {author} {\bibfnamefont {V.}~\bibnamefont
  {Balasubramanian}}\ and\ \bibinfo {author} {\bibfnamefont {P.}~\bibnamefont
  {Kraus}},\ }\href {https://doi.org/10.1103/PhysRevLett.83.3605} {\bibfield
  {journal} {\bibinfo  {journal} {Phys. Rev. Lett.}\ }\textbf {\bibinfo
  {volume} {83}},\ \bibinfo {pages} {3605} (\bibinfo {year} {1999})},\ \Eprint
  {https://arxiv.org/abs/hep-th/9903190} {arXiv:hep-th/9903190} \BibitemShut
  {NoStop}%
\bibitem [{\citenamefont {de~Boer}\ \emph {et~al.}(2000)\citenamefont
  {de~Boer}, \citenamefont {Verlinde},\ and\ \citenamefont
  {Verlinde}}]{deBoer:1999tgo}%
  \BibitemOpen
  \bibfield  {author} {\bibinfo {author} {\bibfnamefont {J.}~\bibnamefont
  {de~Boer}}, \bibinfo {author} {\bibfnamefont {E.~P.}\ \bibnamefont
  {Verlinde}},\ and\ \bibinfo {author} {\bibfnamefont {H.~L.}\ \bibnamefont
  {Verlinde}},\ }\href {https://doi.org/10.1088/1126-6708/2000/08/003}
  {\bibfield  {journal} {\bibinfo  {journal} {JHEP}\ }\textbf {\bibinfo
  {volume} {08}},\ \bibinfo {pages} {003}},\ \Eprint
  {https://arxiv.org/abs/hep-th/9912012} {arXiv:hep-th/9912012} \BibitemShut
  {NoStop}%
\bibitem [{\citenamefont {de~Boer}(2001)}]{deBoer:2000cz}%
  \BibitemOpen
  \bibfield  {author} {\bibinfo {author} {\bibfnamefont {J.}~\bibnamefont
  {de~Boer}},\ }\href
  {https://doi.org/10.1002/1521-3978(200105)49:4/6<339::AID-PROP339>3.0.CO;2-A}
  {\bibfield  {journal} {\bibinfo  {journal} {Fortsch. Phys.}\ }\textbf
  {\bibinfo {volume} {49}},\ \bibinfo {pages} {339} (\bibinfo {year} {2001})},\
  \Eprint {https://arxiv.org/abs/hep-th/0101026} {arXiv:hep-th/0101026}
  \BibitemShut {NoStop}%
\bibitem [{\citenamefont {Bianchi}\ \emph {et~al.}(2002)\citenamefont
  {Bianchi}, \citenamefont {Freedman},\ and\ \citenamefont
  {Skenderis}}]{Bianchi:2001kw}%
  \BibitemOpen
  \bibfield  {author} {\bibinfo {author} {\bibfnamefont {M.}~\bibnamefont
  {Bianchi}}, \bibinfo {author} {\bibfnamefont {D.~Z.}\ \bibnamefont
  {Freedman}},\ and\ \bibinfo {author} {\bibfnamefont {K.}~\bibnamefont
  {Skenderis}},\ }\href {https://doi.org/10.1016/S0550-3213(02)00179-7}
  {\bibfield  {journal} {\bibinfo  {journal} {Nucl. Phys. B}\ }\textbf
  {\bibinfo {volume} {631}},\ \bibinfo {pages} {159} (\bibinfo {year}
  {2002})},\ \Eprint {https://arxiv.org/abs/hep-th/0112119}
  {arXiv:hep-th/0112119} \BibitemShut {NoStop}%
\bibitem [{\citenamefont {Fukuma}\ \emph {et~al.}(2003)\citenamefont {Fukuma},
  \citenamefont {Matsuura},\ and\ \citenamefont {Sakai}}]{Fukuma:2002sb}%
  \BibitemOpen
  \bibfield  {author} {\bibinfo {author} {\bibfnamefont {M.}~\bibnamefont
  {Fukuma}}, \bibinfo {author} {\bibfnamefont {S.}~\bibnamefont {Matsuura}},\
  and\ \bibinfo {author} {\bibfnamefont {T.}~\bibnamefont {Sakai}},\ }\href
  {https://doi.org/10.1143/PTP.109.489} {\bibfield  {journal} {\bibinfo
  {journal} {Prog. Theor. Phys.}\ }\textbf {\bibinfo {volume} {109}},\ \bibinfo
  {pages} {489} (\bibinfo {year} {2003})},\ \Eprint
  {https://arxiv.org/abs/hep-th/0212314} {arXiv:hep-th/0212314} \BibitemShut
  {NoStop}%
\bibitem [{\citenamefont {Papadimitriou}\ and\ \citenamefont
  {Skenderis}(2005{\natexlab{a}})}]{Papadimitriou:2004ap}%
  \BibitemOpen
  \bibfield  {author} {\bibinfo {author} {\bibfnamefont {I.}~\bibnamefont
  {Papadimitriou}}\ and\ \bibinfo {author} {\bibfnamefont {K.}~\bibnamefont
  {Skenderis}},\ }\href {https://doi.org/10.4171/013-1/4} {\bibfield  {journal}
  {\bibinfo  {journal} {IRMA Lect. Math. Theor. Phys.}\ }\textbf {\bibinfo
  {volume} {8}},\ \bibinfo {pages} {73} (\bibinfo {year}
  {2005}{\natexlab{a}})},\ \Eprint {https://arxiv.org/abs/hep-th/0404176}
  {arXiv:hep-th/0404176} \BibitemShut {NoStop}%
\bibitem [{\citenamefont {Papadimitriou}\ and\ \citenamefont
  {Skenderis}(2005{\natexlab{b}})}]{Papadimitriou:2005ii}%
  \BibitemOpen
  \bibfield  {author} {\bibinfo {author} {\bibfnamefont {I.}~\bibnamefont
  {Papadimitriou}}\ and\ \bibinfo {author} {\bibfnamefont {K.}~\bibnamefont
  {Skenderis}},\ }\href {https://doi.org/10.1088/1126-6708/2005/08/004}
  {\bibfield  {journal} {\bibinfo  {journal} {JHEP}\ }\textbf {\bibinfo
  {volume} {08}},\ \bibinfo {pages} {004}},\ \Eprint
  {https://arxiv.org/abs/hep-th/0505190} {arXiv:hep-th/0505190} \BibitemShut
  {NoStop}%
\bibitem [{\citenamefont {Aharony}\ \emph {et~al.}(2000)\citenamefont
  {Aharony}, \citenamefont {Gubser}, \citenamefont {Maldacena}, \citenamefont
  {Ooguri},\ and\ \citenamefont {Oz}}]{Aharony:1999ti}%
  \BibitemOpen
  \bibfield  {author} {\bibinfo {author} {\bibfnamefont {O.}~\bibnamefont
  {Aharony}}, \bibinfo {author} {\bibfnamefont {S.~S.}\ \bibnamefont {Gubser}},
  \bibinfo {author} {\bibfnamefont {J.~M.}\ \bibnamefont {Maldacena}}, \bibinfo
  {author} {\bibfnamefont {H.}~\bibnamefont {Ooguri}},\ and\ \bibinfo {author}
  {\bibfnamefont {Y.}~\bibnamefont {Oz}},\ }\href
  {https://doi.org/10.1016/S0370-1573(99)00083-6} {\bibfield  {journal}
  {\bibinfo  {journal} {Phys. Rept.}\ }\textbf {\bibinfo {volume} {323}},\
  \bibinfo {pages} {183} (\bibinfo {year} {2000})},\ \Eprint
  {https://arxiv.org/abs/hep-th/9905111} {arXiv:hep-th/9905111} \BibitemShut
  {NoStop}%
\bibitem [{\citenamefont {Bourdier}\ and\ \citenamefont
  {Kiritsis}(2014)}]{Bourdier:2013axa}%
  \BibitemOpen
  \bibfield  {author} {\bibinfo {author} {\bibfnamefont {J.}~\bibnamefont
  {Bourdier}}\ and\ \bibinfo {author} {\bibfnamefont {E.}~\bibnamefont
  {Kiritsis}},\ }\href {https://doi.org/10.1088/0264-9381/31/3/035011}
  {\bibfield  {journal} {\bibinfo  {journal} {Class. Quant. Grav.}\ }\textbf
  {\bibinfo {volume} {31}},\ \bibinfo {pages} {035011} (\bibinfo {year}
  {2014})},\ \Eprint {https://arxiv.org/abs/1310.0858} {arXiv:1310.0858
  [hep-th]} \BibitemShut {NoStop}%
\bibitem [{\citenamefont {Kiritsis}\ \emph {et~al.}(2014)\citenamefont
  {Kiritsis}, \citenamefont {Li},\ and\ \citenamefont
  {Nitti}}]{Kiritsis:2014kua}%
  \BibitemOpen
  \bibfield  {author} {\bibinfo {author} {\bibfnamefont {E.}~\bibnamefont
  {Kiritsis}}, \bibinfo {author} {\bibfnamefont {W.}~\bibnamefont {Li}},\ and\
  \bibinfo {author} {\bibfnamefont {F.}~\bibnamefont {Nitti}},\ }\href
  {https://doi.org/10.1002/prop.201400007} {\bibfield  {journal} {\bibinfo
  {journal} {Fortsch. Phys.}\ }\textbf {\bibinfo {volume} {62}},\ \bibinfo
  {pages} {389} (\bibinfo {year} {2014})},\ \Eprint
  {https://arxiv.org/abs/1401.0888} {arXiv:1401.0888 [hep-th]} \BibitemShut
  {NoStop}%
\bibitem [{\citenamefont {Kiritsis}\ \emph {et~al.}(2017)\citenamefont
  {Kiritsis}, \citenamefont {Nitti},\ and\ \citenamefont
  {Silva~Pimenta}}]{Kiritsis:2016kog}%
  \BibitemOpen
  \bibfield  {author} {\bibinfo {author} {\bibfnamefont {E.}~\bibnamefont
  {Kiritsis}}, \bibinfo {author} {\bibfnamefont {F.}~\bibnamefont {Nitti}},\
  and\ \bibinfo {author} {\bibfnamefont {L.}~\bibnamefont {Silva~Pimenta}},\
  }\href {https://doi.org/10.1002/prop.201600120} {\bibfield  {journal}
  {\bibinfo  {journal} {Fortsch. Phys.}\ }\textbf {\bibinfo {volume} {65}},\
  \bibinfo {pages} {1600120} (\bibinfo {year} {2017})},\ \Eprint
  {https://arxiv.org/abs/1611.05493} {arXiv:1611.05493 [hep-th]} \BibitemShut
  {NoStop}%
\bibitem [{\citenamefont {G\"ursoy}\ \emph {et~al.}(2018)\citenamefont
  {G\"ursoy}, \citenamefont {Kiritsis}, \citenamefont {Nitti},\ and\
  \citenamefont {Silva~Pimenta}}]{Gursoy:2018umf}%
  \BibitemOpen
  \bibfield  {author} {\bibinfo {author} {\bibfnamefont {U.}~\bibnamefont
  {G\"ursoy}}, \bibinfo {author} {\bibfnamefont {E.}~\bibnamefont {Kiritsis}},
  \bibinfo {author} {\bibfnamefont {F.}~\bibnamefont {Nitti}},\ and\ \bibinfo
  {author} {\bibfnamefont {L.}~\bibnamefont {Silva~Pimenta}},\ }\href
  {https://doi.org/10.1007/JHEP10(2018)173} {\bibfield  {journal} {\bibinfo
  {journal} {JHEP}\ }\textbf {\bibinfo {volume} {10}},\ \bibinfo {pages}
  {173}},\ \Eprint {https://arxiv.org/abs/1805.01769} {arXiv:1805.01769
  [hep-th]} \BibitemShut {NoStop}%
\bibitem [{\citenamefont {Kasner}(1921)}]{Kasner:1921zz}%
  \BibitemOpen
  \bibfield  {author} {\bibinfo {author} {\bibfnamefont {E.}~\bibnamefont
  {Kasner}},\ }\href {https://doi.org/10.2307/2370192} {\bibfield  {journal}
  {\bibinfo  {journal} {Am. J. Math.}\ }\textbf {\bibinfo {volume} {43}},\
  \bibinfo {pages} {217} (\bibinfo {year} {1921})}\BibitemShut {NoStop}%
\bibitem [{\citenamefont {Belinskii}\ and\ \citenamefont
  {Khalatnikov}(1973)}]{Belinskii:1973zz}%
  \BibitemOpen
  \bibfield  {author} {\bibinfo {author} {\bibfnamefont {V.~A.}\ \bibnamefont
  {Belinskii}}\ and\ \bibinfo {author} {\bibfnamefont {I.~M.}\ \bibnamefont
  {Khalatnikov}},\ }\href@noop {} {\bibfield  {journal} {\bibinfo  {journal}
  {Sov. Phys. JETP}\ }\textbf {\bibinfo {volume} {36}},\ \bibinfo {pages} {591}
  (\bibinfo {year} {1973})}\BibitemShut {NoStop}%
\bibitem [{\citenamefont {Das}\ \emph {et~al.}(2006)\citenamefont {Das},
  \citenamefont {Michelson}, \citenamefont {Narayan},\ and\ \citenamefont
  {Trivedi}}]{Das:2006dz}%
  \BibitemOpen
  \bibfield  {author} {\bibinfo {author} {\bibfnamefont {S.~R.}\ \bibnamefont
  {Das}}, \bibinfo {author} {\bibfnamefont {J.}~\bibnamefont {Michelson}},
  \bibinfo {author} {\bibfnamefont {K.}~\bibnamefont {Narayan}},\ and\ \bibinfo
  {author} {\bibfnamefont {S.~P.}\ \bibnamefont {Trivedi}},\ }\href
  {https://doi.org/10.1103/PhysRevD.74.026002} {\bibfield  {journal} {\bibinfo
  {journal} {Phys. Rev. D}\ }\textbf {\bibinfo {volume} {74}},\ \bibinfo
  {pages} {026002} (\bibinfo {year} {2006})},\ \Eprint
  {https://arxiv.org/abs/hep-th/0602107} {arXiv:hep-th/0602107} \BibitemShut
  {NoStop}%
\bibitem [{\citenamefont {Hartman}\ and\ \citenamefont
  {Maldacena}(2013)}]{Hartman:2013qma}%
  \BibitemOpen
  \bibfield  {author} {\bibinfo {author} {\bibfnamefont {T.}~\bibnamefont
  {Hartman}}\ and\ \bibinfo {author} {\bibfnamefont {J.}~\bibnamefont
  {Maldacena}},\ }\href {https://doi.org/10.1007/JHEP05(2013)014} {\bibfield
  {journal} {\bibinfo  {journal} {JHEP}\ }\textbf {\bibinfo {volume} {05}},\
  \bibinfo {pages} {014}},\ \Eprint {https://arxiv.org/abs/1303.1080}
  {arXiv:1303.1080 [hep-th]} \BibitemShut {NoStop}%
\bibitem [{\citenamefont {Wang}\ \emph {et~al.}(2020)\citenamefont {Wang},
  \citenamefont {Song}, \citenamefont {Xiang}, \citenamefont {Wei},
  \citenamefont {Zhu},\ and\ \citenamefont {Liu}}]{Wang:2020nkd}%
  \BibitemOpen
  \bibfield  {author} {\bibinfo {author} {\bibfnamefont {Y.-Q.}\ \bibnamefont
  {Wang}}, \bibinfo {author} {\bibfnamefont {Y.}~\bibnamefont {Song}}, \bibinfo
  {author} {\bibfnamefont {Q.}~\bibnamefont {Xiang}}, \bibinfo {author}
  {\bibfnamefont {S.-W.}\ \bibnamefont {Wei}}, \bibinfo {author} {\bibfnamefont
  {T.}~\bibnamefont {Zhu}},\ and\ \bibinfo {author} {\bibfnamefont {Y.-X.}\
  \bibnamefont {Liu}},\ }\href@noop {} {\  (\bibinfo {year} {2020})},\ \Eprint
  {https://arxiv.org/abs/2009.06277} {arXiv:2009.06277 [hep-th]} \BibitemShut
  {NoStop}%
\bibitem [{\citenamefont {Hartnoll}\ \emph {et~al.}(2021)\citenamefont
  {Hartnoll}, \citenamefont {Horowitz}, \citenamefont {Kruthoff},\ and\
  \citenamefont {Santos}}]{Hartnoll:2020fhc}%
  \BibitemOpen
  \bibfield  {author} {\bibinfo {author} {\bibfnamefont {S.~A.}\ \bibnamefont
  {Hartnoll}}, \bibinfo {author} {\bibfnamefont {G.~T.}\ \bibnamefont
  {Horowitz}}, \bibinfo {author} {\bibfnamefont {J.}~\bibnamefont {Kruthoff}},\
  and\ \bibinfo {author} {\bibfnamefont {J.~E.}\ \bibnamefont {Santos}},\
  }\href {https://doi.org/10.21468/SciPostPhys.10.1.009} {\bibfield  {journal}
  {\bibinfo  {journal} {SciPost Phys.}\ }\textbf {\bibinfo {volume} {10}},\
  \bibinfo {pages} {009} (\bibinfo {year} {2021})},\ \Eprint
  {https://arxiv.org/abs/2008.12786} {arXiv:2008.12786 [hep-th]} \BibitemShut
  {NoStop}%
\bibitem [{\citenamefont {Sword}\ and\ \citenamefont
  {Vegh}(2021)}]{Sword:2021pfm}%
  \BibitemOpen
  \bibfield  {author} {\bibinfo {author} {\bibfnamefont {L.}~\bibnamefont
  {Sword}}\ and\ \bibinfo {author} {\bibfnamefont {D.}~\bibnamefont {Vegh}},\
  }\href@noop {} {\  (\bibinfo {year} {2021})},\ \Eprint
  {https://arxiv.org/abs/2112.14177} {arXiv:2112.14177 [hep-th]} \BibitemShut
  {NoStop}%
\bibitem [{\citenamefont {Mansoori}\ \emph {et~al.}(2021)\citenamefont
  {Mansoori}, \citenamefont {Li}, \citenamefont {Rafiee},\ and\ \citenamefont
  {Baggioli}}]{Mansoori:2021wxf}%
  \BibitemOpen
  \bibfield  {author} {\bibinfo {author} {\bibfnamefont {S.~A.~H.}\
  \bibnamefont {Mansoori}}, \bibinfo {author} {\bibfnamefont {L.}~\bibnamefont
  {Li}}, \bibinfo {author} {\bibfnamefont {M.}~\bibnamefont {Rafiee}},\ and\
  \bibinfo {author} {\bibfnamefont {M.}~\bibnamefont {Baggioli}},\ }\href
  {https://doi.org/10.1007/JHEP10(2021)098} {\bibfield  {journal} {\bibinfo
  {journal} {JHEP}\ }\textbf {\bibinfo {volume} {10}},\ \bibinfo {pages}
  {098}},\ \Eprint {https://arxiv.org/abs/2108.01471} {arXiv:2108.01471
  [hep-th]} \BibitemShut {NoStop}%
\bibitem [{\citenamefont {Das}\ and\ \citenamefont
  {Kundu}(2021)}]{Das:2021vjf}%
  \BibitemOpen
  \bibfield  {author} {\bibinfo {author} {\bibfnamefont {S.}~\bibnamefont
  {Das}}\ and\ \bibinfo {author} {\bibfnamefont {A.}~\bibnamefont {Kundu}},\
  }\href@noop {} {\  (\bibinfo {year} {2021})},\ \Eprint
  {https://arxiv.org/abs/2112.11675} {arXiv:2112.11675 [hep-th]} \BibitemShut
  {NoStop}%
\bibitem [{\citenamefont {Lifshitz}\ and\ \citenamefont
  {Khalatnikov}(1963)}]{Lifshitz:1963ps}%
  \BibitemOpen
  \bibfield  {author} {\bibinfo {author} {\bibfnamefont {E.~M.}\ \bibnamefont
  {Lifshitz}}\ and\ \bibinfo {author} {\bibfnamefont {I.~M.}\ \bibnamefont
  {Khalatnikov}},\ }\href {https://doi.org/10.1080/00018736300101283}
  {\bibfield  {journal} {\bibinfo  {journal} {Adv. Phys.}\ }\textbf {\bibinfo
  {volume} {12}},\ \bibinfo {pages} {185} (\bibinfo {year} {1963})}\BibitemShut
  {NoStop}%
\bibitem [{\citenamefont {Belinskii}\ \emph {et~al.}(1970)\citenamefont
  {Belinskii}, \citenamefont {Khalatnikov},\ and\ \citenamefont
  {Lifshitz}}]{Belinskii:1970ew}%
  \BibitemOpen
  \bibfield  {author} {\bibinfo {author} {\bibfnamefont {V.~A.}\ \bibnamefont
  {Belinskii}}, \bibinfo {author} {\bibfnamefont {I.~M.}\ \bibnamefont
  {Khalatnikov}},\ and\ \bibinfo {author} {\bibfnamefont {E.~M.}\ \bibnamefont
  {Lifshitz}},\ }\href {https://doi.org/10.1080/00018737000101171} {\bibfield
  {journal} {\bibinfo  {journal} {Adv. Phys.}\ }\textbf {\bibinfo {volume}
  {19}},\ \bibinfo {pages} {525} (\bibinfo {year} {1970})}\BibitemShut
  {NoStop}%
\bibitem [{\citenamefont {Belinskii}\ \emph {et~al.}(1982)\citenamefont
  {Belinskii}, \citenamefont {Khalatnikov},\ and\ \citenamefont
  {Lifshitz}}]{Belinskii:1982pk}%
  \BibitemOpen
  \bibfield  {author} {\bibinfo {author} {\bibfnamefont {V.~A.}\ \bibnamefont
  {Belinskii}}, \bibinfo {author} {\bibfnamefont {I.~M.}\ \bibnamefont
  {Khalatnikov}},\ and\ \bibinfo {author} {\bibfnamefont {E.~M.}\ \bibnamefont
  {Lifshitz}},\ }\href {https://doi.org/10.1080/00018738200101428} {\bibfield
  {journal} {\bibinfo  {journal} {Adv. Phys.}\ }\textbf {\bibinfo {volume}
  {31}},\ \bibinfo {pages} {639} (\bibinfo {year} {1982})}\BibitemShut
  {NoStop}%
\bibitem [{\citenamefont {Belinskii}(2010)}]{Belinskii:2009wj}%
  \BibitemOpen
  \bibfield  {author} {\bibinfo {author} {\bibfnamefont {V.~A.}\ \bibnamefont
  {Belinskii}},\ }\href {https://doi.org/10.1063/1.3382327} {\bibfield
  {journal} {\bibinfo  {journal} {AIP Conf. Proc.}\ }\textbf {\bibinfo {volume}
  {1205}},\ \bibinfo {pages} {17} (\bibinfo {year} {2010})},\ \Eprint
  {https://arxiv.org/abs/0910.0374} {arXiv:0910.0374 [gr-qc]} \BibitemShut
  {NoStop}%
\bibitem [{\citenamefont {Misner}(1969)}]{Misner:1969hg}%
  \BibitemOpen
  \bibfield  {author} {\bibinfo {author} {\bibfnamefont {C.~W.}\ \bibnamefont
  {Misner}},\ }\href {https://doi.org/10.1103/PhysRevLett.22.1071} {\bibfield
  {journal} {\bibinfo  {journal} {Phys. Rev. Lett.}\ }\textbf {\bibinfo
  {volume} {22}},\ \bibinfo {pages} {1071} (\bibinfo {year}
  {1969})}\BibitemShut {NoStop}%
\bibitem [{\citenamefont {Damour}\ \emph {et~al.}(2003)\citenamefont {Damour},
  \citenamefont {Henneaux},\ and\ \citenamefont {Nicolai}}]{Damour:2002et}%
  \BibitemOpen
  \bibfield  {author} {\bibinfo {author} {\bibfnamefont {T.}~\bibnamefont
  {Damour}}, \bibinfo {author} {\bibfnamefont {M.}~\bibnamefont {Henneaux}},\
  and\ \bibinfo {author} {\bibfnamefont {H.}~\bibnamefont {Nicolai}},\ }\href
  {https://doi.org/10.1088/0264-9381/20/9/201} {\bibfield  {journal} {\bibinfo
  {journal} {Class. Quant. Grav.}\ }\textbf {\bibinfo {volume} {20}},\ \bibinfo
  {pages} {R145} (\bibinfo {year} {2003})},\ \Eprint
  {https://arxiv.org/abs/hep-th/0212256} {arXiv:hep-th/0212256} \BibitemShut
  {NoStop}%
\bibitem [{\citenamefont {Damour}\ \emph {et~al.}(2002)\citenamefont {Damour},
  \citenamefont {Henneaux}, \citenamefont {Rendall},\ and\ \citenamefont
  {Weaver}}]{Damour:2002tc}%
  \BibitemOpen
  \bibfield  {author} {\bibinfo {author} {\bibfnamefont {T.}~\bibnamefont
  {Damour}}, \bibinfo {author} {\bibfnamefont {M.}~\bibnamefont {Henneaux}},
  \bibinfo {author} {\bibfnamefont {A.~D.}\ \bibnamefont {Rendall}},\ and\
  \bibinfo {author} {\bibfnamefont {M.}~\bibnamefont {Weaver}},\ }\href
  {https://doi.org/10.1007/s000230200000} {\bibfield  {journal} {\bibinfo
  {journal} {Annales Henri Poincare}\ }\textbf {\bibinfo {volume} {3}},\
  \bibinfo {pages} {1049} (\bibinfo {year} {2002})},\ \Eprint
  {https://arxiv.org/abs/gr-qc/0202069} {arXiv:gr-qc/0202069} \BibitemShut
  {NoStop}%
\bibitem [{\citenamefont {Klebanov}\ and\ \citenamefont
  {Witten}(1999)}]{Klebanov:1999tb}%
  \BibitemOpen
  \bibfield  {author} {\bibinfo {author} {\bibfnamefont {I.~R.}\ \bibnamefont
  {Klebanov}}\ and\ \bibinfo {author} {\bibfnamefont {E.}~\bibnamefont
  {Witten}},\ }\href {https://doi.org/10.1016/S0550-3213(99)00387-9} {\bibfield
   {journal} {\bibinfo  {journal} {Nucl. Phys. B}\ }\textbf {\bibinfo {volume}
  {556}},\ \bibinfo {pages} {89} (\bibinfo {year} {1999})},\ \Eprint
  {https://arxiv.org/abs/hep-th/9905104} {arXiv:hep-th/9905104} \BibitemShut
  {NoStop}%
\bibitem [{\citenamefont {Minces}\ and\ \citenamefont
  {Rivelles}(2000)}]{Minces:1999eg}%
  \BibitemOpen
  \bibfield  {author} {\bibinfo {author} {\bibfnamefont {P.}~\bibnamefont
  {Minces}}\ and\ \bibinfo {author} {\bibfnamefont {V.~O.}\ \bibnamefont
  {Rivelles}},\ }\href {https://doi.org/10.1016/S0550-3213(99)00833-0}
  {\bibfield  {journal} {\bibinfo  {journal} {Nucl. Phys. B}\ }\textbf
  {\bibinfo {volume} {572}},\ \bibinfo {pages} {651} (\bibinfo {year}
  {2000})},\ \Eprint {https://arxiv.org/abs/hep-th/9907079}
  {arXiv:hep-th/9907079} \BibitemShut {NoStop}%
\bibitem [{\citenamefont {Doroshkevich}\ and\ \citenamefont
  {Novikov}(1978)}]{Doroshkevich:1978aq}%
  \BibitemOpen
  \bibfield  {author} {\bibinfo {author} {\bibfnamefont {A.~G.}\ \bibnamefont
  {Doroshkevich}}\ and\ \bibinfo {author} {\bibfnamefont {I.~D.}\ \bibnamefont
  {Novikov}},\ }\href@noop {} {\bibfield  {journal} {\bibinfo  {journal} {Zh.
  Eksp. Teor. Fiz.}\ }\textbf {\bibinfo {volume} {74}},\ \bibinfo {pages} {3}
  (\bibinfo {year} {1978})}\BibitemShut {NoStop}%
\bibitem [{\citenamefont {Fournodavlos}\ and\ \citenamefont
  {Sbierski}(2020)}]{Fournodavlos:2018lrk}%
  \BibitemOpen
  \bibfield  {author} {\bibinfo {author} {\bibfnamefont {G.}~\bibnamefont
  {Fournodavlos}}\ and\ \bibinfo {author} {\bibfnamefont {J.}~\bibnamefont
  {Sbierski}},\ }\href {https://doi.org/10.1007/s00205-019-01434-0} {\bibfield
  {journal} {\bibinfo  {journal} {Arch. Ration. Mech. Anal.}\ }\textbf
  {\bibinfo {volume} {235}},\ \bibinfo {pages} {927} (\bibinfo {year}
  {2020})},\ \Eprint {https://arxiv.org/abs/1804.01941} {arXiv:1804.01941
  [gr-qc]} \BibitemShut {NoStop}%
\bibitem [{\citenamefont {Grandi}\ and\ \citenamefont
  {Salazar~Landea}(2021)}]{Grandi:2021ajl}%
  \BibitemOpen
  \bibfield  {author} {\bibinfo {author} {\bibfnamefont {N.}~\bibnamefont
  {Grandi}}\ and\ \bibinfo {author} {\bibfnamefont {I.}~\bibnamefont
  {Salazar~Landea}},\ }\href {https://doi.org/10.1007/JHEP05(2021)152}
  {\bibfield  {journal} {\bibinfo  {journal} {JHEP}\ }\textbf {\bibinfo
  {volume} {05}},\ \bibinfo {pages} {152}},\ \Eprint
  {https://arxiv.org/abs/2102.02707} {arXiv:2102.02707 [gr-qc]} \BibitemShut
  {NoStop}%
\bibitem [{\citenamefont {Hartnoll}\ \emph {et~al.}(2020)\citenamefont
  {Hartnoll}, \citenamefont {Horowitz}, \citenamefont {Kruthoff},\ and\
  \citenamefont {Santos}}]{Hartnoll:2020rwq}%
  \BibitemOpen
  \bibfield  {author} {\bibinfo {author} {\bibfnamefont {S.~A.}\ \bibnamefont
  {Hartnoll}}, \bibinfo {author} {\bibfnamefont {G.~T.}\ \bibnamefont
  {Horowitz}}, \bibinfo {author} {\bibfnamefont {J.}~\bibnamefont {Kruthoff}},\
  and\ \bibinfo {author} {\bibfnamefont {J.~E.}\ \bibnamefont {Santos}},\
  }\href {https://doi.org/10.1007/JHEP10(2020)102} {\bibfield  {journal}
  {\bibinfo  {journal} {JHEP}\ }\textbf {\bibinfo {volume} {10}},\ \bibinfo
  {pages} {102}},\ \Eprint {https://arxiv.org/abs/2006.10056} {arXiv:2006.10056
  [hep-th]} \BibitemShut {NoStop}%
\bibitem [{\citenamefont {Cai}\ \emph {et~al.}(2021{\natexlab{a}})\citenamefont
  {Cai}, \citenamefont {Li},\ and\ \citenamefont {Yang}}]{Cai:2020wrp}%
  \BibitemOpen
  \bibfield  {author} {\bibinfo {author} {\bibfnamefont {R.-G.}\ \bibnamefont
  {Cai}}, \bibinfo {author} {\bibfnamefont {L.}~\bibnamefont {Li}},\ and\
  \bibinfo {author} {\bibfnamefont {R.-Q.}\ \bibnamefont {Yang}},\ }\href
  {https://doi.org/10.1007/JHEP03(2021)263} {\bibfield  {journal} {\bibinfo
  {journal} {JHEP}\ }\textbf {\bibinfo {volume} {03}},\ \bibinfo {pages}
  {263}},\ \Eprint {https://arxiv.org/abs/2009.05520} {arXiv:2009.05520
  [gr-qc]} \BibitemShut {NoStop}%
\bibitem [{\citenamefont {Cai}\ \emph {et~al.}(2021{\natexlab{b}})\citenamefont
  {Cai}, \citenamefont {Ge}, \citenamefont {Li},\ and\ \citenamefont
  {Yang}}]{Cai:2021obq}%
  \BibitemOpen
  \bibfield  {author} {\bibinfo {author} {\bibfnamefont {R.-G.}\ \bibnamefont
  {Cai}}, \bibinfo {author} {\bibfnamefont {C.}~\bibnamefont {Ge}}, \bibinfo
  {author} {\bibfnamefont {L.}~\bibnamefont {Li}},\ and\ \bibinfo {author}
  {\bibfnamefont {R.-Q.}\ \bibnamefont {Yang}},\ }\href@noop {} {\  (\bibinfo
  {year} {2021}{\natexlab{b}})},\ \Eprint {https://arxiv.org/abs/2112.04206}
  {arXiv:2112.04206 [gr-qc]} \BibitemShut {NoStop}%
\bibitem [{\citenamefont {Carmi}\ \emph {et~al.}(2017)\citenamefont {Carmi},
  \citenamefont {Chapman}, \citenamefont {Marrochio}, \citenamefont {Myers},\
  and\ \citenamefont {Sugishita}}]{Carmi:2017jqz}%
  \BibitemOpen
  \bibfield  {author} {\bibinfo {author} {\bibfnamefont {D.}~\bibnamefont
  {Carmi}}, \bibinfo {author} {\bibfnamefont {S.}~\bibnamefont {Chapman}},
  \bibinfo {author} {\bibfnamefont {H.}~\bibnamefont {Marrochio}}, \bibinfo
  {author} {\bibfnamefont {R.~C.}\ \bibnamefont {Myers}},\ and\ \bibinfo
  {author} {\bibfnamefont {S.}~\bibnamefont {Sugishita}},\ }\href
  {https://doi.org/10.1007/JHEP11(2017)188} {\bibfield  {journal} {\bibinfo
  {journal} {JHEP}\ }\textbf {\bibinfo {volume} {11}},\ \bibinfo {pages}
  {188}},\ \Eprint {https://arxiv.org/abs/1709.10184} {arXiv:1709.10184
  [hep-th]} \BibitemShut {NoStop}%
\bibitem [{\citenamefont {Wall}(2014)}]{Wall:2012uf}%
  \BibitemOpen
  \bibfield  {author} {\bibinfo {author} {\bibfnamefont {A.~C.}\ \bibnamefont
  {Wall}},\ }\href {https://doi.org/10.1088/0264-9381/31/22/225007} {\bibfield
  {journal} {\bibinfo  {journal} {Class. Quant. Grav.}\ }\textbf {\bibinfo
  {volume} {31}},\ \bibinfo {pages} {225007} (\bibinfo {year} {2014})},\
  \Eprint {https://arxiv.org/abs/1211.3494} {arXiv:1211.3494 [hep-th]}
  \BibitemShut {NoStop}%
\bibitem [{\citenamefont {Engelhardt}\ and\ \citenamefont
  {Wall}(2014)}]{Engelhardt:2013tra}%
  \BibitemOpen
  \bibfield  {author} {\bibinfo {author} {\bibfnamefont {N.}~\bibnamefont
  {Engelhardt}}\ and\ \bibinfo {author} {\bibfnamefont {A.~C.}\ \bibnamefont
  {Wall}},\ }\href {https://doi.org/10.1007/JHEP03(2014)068} {\bibfield
  {journal} {\bibinfo  {journal} {JHEP}\ }\textbf {\bibinfo {volume} {03}},\
  \bibinfo {pages} {068}},\ \Eprint {https://arxiv.org/abs/1312.3699}
  {arXiv:1312.3699 [hep-th]} \BibitemShut {NoStop}%
\bibitem [{\citenamefont {Casini}\ \emph
  {et~al.}(2017{\natexlab{a}})\citenamefont {Casini}, \citenamefont {Teste},\
  and\ \citenamefont {Torroba}}]{Casini:2016udt}%
  \BibitemOpen
  \bibfield  {author} {\bibinfo {author} {\bibfnamefont {H.}~\bibnamefont
  {Casini}}, \bibinfo {author} {\bibfnamefont {E.}~\bibnamefont {Teste}},\ and\
  \bibinfo {author} {\bibfnamefont {G.}~\bibnamefont {Torroba}},\ }\href
  {https://doi.org/10.1007/JHEP03(2017)089} {\bibfield  {journal} {\bibinfo
  {journal} {JHEP}\ }\textbf {\bibinfo {volume} {03}},\ \bibinfo {pages}
  {089}},\ \Eprint {https://arxiv.org/abs/1611.00016} {arXiv:1611.00016
  [hep-th]} \BibitemShut {NoStop}%
\bibitem [{\citenamefont {Casini}\ \emph
  {et~al.}(2017{\natexlab{b}})\citenamefont {Casini}, \citenamefont {Test\'e},\
  and\ \citenamefont {Torroba}}]{Casini:2017vbe}%
  \BibitemOpen
  \bibfield  {author} {\bibinfo {author} {\bibfnamefont {H.}~\bibnamefont
  {Casini}}, \bibinfo {author} {\bibfnamefont {E.}~\bibnamefont {Test\'e}},\
  and\ \bibinfo {author} {\bibfnamefont {G.}~\bibnamefont {Torroba}},\ }\href
  {https://doi.org/10.1103/PhysRevLett.118.261602} {\bibfield  {journal}
  {\bibinfo  {journal} {Phys. Rev. Lett.}\ }\textbf {\bibinfo {volume} {118}},\
  \bibinfo {pages} {261602} (\bibinfo {year} {2017}{\natexlab{b}})},\ \Eprint
  {https://arxiv.org/abs/1704.01870} {arXiv:1704.01870 [hep-th]} \BibitemShut
  {NoStop}%
\bibitem [{\citenamefont {Akers}\ \emph {et~al.}(2016)\citenamefont {Akers},
  \citenamefont {Koeller}, \citenamefont {Leichenauer},\ and\ \citenamefont
  {Levine}}]{Akers:2016ugt}%
  \BibitemOpen
  \bibfield  {author} {\bibinfo {author} {\bibfnamefont {C.}~\bibnamefont
  {Akers}}, \bibinfo {author} {\bibfnamefont {J.}~\bibnamefont {Koeller}},
  \bibinfo {author} {\bibfnamefont {S.}~\bibnamefont {Leichenauer}},\ and\
  \bibinfo {author} {\bibfnamefont {A.}~\bibnamefont {Levine}},\ }\href@noop {}
  {\  (\bibinfo {year} {2016})},\ \Eprint {https://arxiv.org/abs/1610.08968}
  {arXiv:1610.08968 [hep-th]} \BibitemShut {NoStop}%
\bibitem [{\citenamefont {Akers}\ \emph {et~al.}(2020)\citenamefont {Akers},
  \citenamefont {Chandrasekaran}, \citenamefont {Leichenauer}, \citenamefont
  {Levine},\ and\ \citenamefont {Shahbazi~Moghaddam}}]{Akers:2017ttv}%
  \BibitemOpen
  \bibfield  {author} {\bibinfo {author} {\bibfnamefont {C.}~\bibnamefont
  {Akers}}, \bibinfo {author} {\bibfnamefont {V.}~\bibnamefont
  {Chandrasekaran}}, \bibinfo {author} {\bibfnamefont {S.}~\bibnamefont
  {Leichenauer}}, \bibinfo {author} {\bibfnamefont {A.}~\bibnamefont
  {Levine}},\ and\ \bibinfo {author} {\bibfnamefont {A.}~\bibnamefont
  {Shahbazi~Moghaddam}},\ }\href {https://doi.org/10.1103/PhysRevD.101.025011}
  {\bibfield  {journal} {\bibinfo  {journal} {Phys. Rev. D}\ }\textbf {\bibinfo
  {volume} {101}},\ \bibinfo {pages} {025011} (\bibinfo {year} {2020})},\
  \Eprint {https://arxiv.org/abs/1706.04183} {arXiv:1706.04183 [hep-th]}
  \BibitemShut {NoStop}%
\bibitem [{\citenamefont {Hashimoto}\ and\ \citenamefont
  {Watanabe}(2021)}]{Hashimoto:2021umd}%
  \BibitemOpen
  \bibfield  {author} {\bibinfo {author} {\bibfnamefont {K.}~\bibnamefont
  {Hashimoto}}\ and\ \bibinfo {author} {\bibfnamefont {R.}~\bibnamefont
  {Watanabe}},\ }\href@noop {} {\  (\bibinfo {year} {2021})},\ \Eprint
  {https://arxiv.org/abs/2103.13186} {arXiv:2103.13186 [hep-th]} \BibitemShut
  {NoStop}%
\bibitem [{\citenamefont {Auzzi}\ \emph {et~al.}(2022)\citenamefont {Auzzi},
  \citenamefont {Bolognesi}, \citenamefont {Rabinovici}, \citenamefont
  {Schaposnik~Massolo},\ and\ \citenamefont {Tallarita}}]{Auzzi:2022bfd}%
  \BibitemOpen
  \bibfield  {author} {\bibinfo {author} {\bibfnamefont {R.}~\bibnamefont
  {Auzzi}}, \bibinfo {author} {\bibfnamefont {S.}~\bibnamefont {Bolognesi}},
  \bibinfo {author} {\bibfnamefont {E.}~\bibnamefont {Rabinovici}}, \bibinfo
  {author} {\bibfnamefont {F.~I.}\ \bibnamefont {Schaposnik~Massolo}},\ and\
  \bibinfo {author} {\bibfnamefont {G.}~\bibnamefont {Tallarita}},\ }\href@noop
  {} {\  (\bibinfo {year} {2022})},\ \Eprint {https://arxiv.org/abs/2205.03365}
  {arXiv:2205.03365 [hep-th]} \BibitemShut {NoStop}%
\bibitem [{\citenamefont {Lehner}\ \emph {et~al.}(2016)\citenamefont {Lehner},
  \citenamefont {Myers}, \citenamefont {Poisson},\ and\ \citenamefont
  {Sorkin}}]{Lehner:2016vdi}%
  \BibitemOpen
  \bibfield  {author} {\bibinfo {author} {\bibfnamefont {L.}~\bibnamefont
  {Lehner}}, \bibinfo {author} {\bibfnamefont {R.~C.}\ \bibnamefont {Myers}},
  \bibinfo {author} {\bibfnamefont {E.}~\bibnamefont {Poisson}},\ and\ \bibinfo
  {author} {\bibfnamefont {R.~D.}\ \bibnamefont {Sorkin}},\ }\href
  {https://doi.org/10.1103/PhysRevD.94.084046} {\bibfield  {journal} {\bibinfo
  {journal} {Phys. Rev. D}\ }\textbf {\bibinfo {volume} {94}},\ \bibinfo
  {pages} {084046} (\bibinfo {year} {2016})},\ \Eprint
  {https://arxiv.org/abs/1609.00207} {arXiv:1609.00207 [hep-th]} \BibitemShut
  {NoStop}%
\bibitem [{\citenamefont {Chapman}\ \emph {et~al.}(2017)\citenamefont
  {Chapman}, \citenamefont {Marrochio},\ and\ \citenamefont
  {Myers}}]{Chapman:2016hwi}%
  \BibitemOpen
  \bibfield  {author} {\bibinfo {author} {\bibfnamefont {S.}~\bibnamefont
  {Chapman}}, \bibinfo {author} {\bibfnamefont {H.}~\bibnamefont {Marrochio}},\
  and\ \bibinfo {author} {\bibfnamefont {R.~C.}\ \bibnamefont {Myers}},\ }\href
  {https://doi.org/10.1007/JHEP01(2017)062} {\bibfield  {journal} {\bibinfo
  {journal} {JHEP}\ }\textbf {\bibinfo {volume} {01}},\ \bibinfo {pages}
  {062}},\ \Eprint {https://arxiv.org/abs/1610.08063} {arXiv:1610.08063
  [hep-th]} \BibitemShut {NoStop}%
\bibitem [{\citenamefont {Auzzi}\ \emph {et~al.}(2018)\citenamefont {Auzzi},
  \citenamefont {Baiguera}, \citenamefont {Grassi}, \citenamefont {Nardelli},\
  and\ \citenamefont {Zenoni}}]{Auzzi:2018pbc}%
  \BibitemOpen
  \bibfield  {author} {\bibinfo {author} {\bibfnamefont {R.}~\bibnamefont
  {Auzzi}}, \bibinfo {author} {\bibfnamefont {S.}~\bibnamefont {Baiguera}},
  \bibinfo {author} {\bibfnamefont {M.}~\bibnamefont {Grassi}}, \bibinfo
  {author} {\bibfnamefont {G.}~\bibnamefont {Nardelli}},\ and\ \bibinfo
  {author} {\bibfnamefont {N.}~\bibnamefont {Zenoni}},\ }\href
  {https://doi.org/10.1007/JHEP09(2018)013} {\bibfield  {journal} {\bibinfo
  {journal} {JHEP}\ }\textbf {\bibinfo {volume} {09}},\ \bibinfo {pages}
  {013}},\ \Eprint {https://arxiv.org/abs/1806.06216} {arXiv:1806.06216
  [hep-th]} \BibitemShut {NoStop}%
\bibitem [{\citenamefont {Bernamonti}\ \emph {et~al.}(2021)\citenamefont
  {Bernamonti}, \citenamefont {Bigazzi}, \citenamefont {Billo}, \citenamefont
  {Faggi},\ and\ \citenamefont {Galli}}]{Bernamonti:2021jyu}%
  \BibitemOpen
  \bibfield  {author} {\bibinfo {author} {\bibfnamefont {A.}~\bibnamefont
  {Bernamonti}}, \bibinfo {author} {\bibfnamefont {F.}~\bibnamefont {Bigazzi}},
  \bibinfo {author} {\bibfnamefont {D.}~\bibnamefont {Billo}}, \bibinfo
  {author} {\bibfnamefont {L.}~\bibnamefont {Faggi}},\ and\ \bibinfo {author}
  {\bibfnamefont {F.}~\bibnamefont {Galli}},\ }\href
  {https://doi.org/10.1007/JHEP11(2021)037} {\bibfield  {journal} {\bibinfo
  {journal} {JHEP}\ }\textbf {\bibinfo {volume} {11}},\ \bibinfo {pages}
  {037}},\ \Eprint {https://arxiv.org/abs/2108.09281} {arXiv:2108.09281
  [hep-th]} \BibitemShut {NoStop}%
\bibitem [{\citenamefont {Belin}\ \emph {et~al.}(2015)\citenamefont {Belin},
  \citenamefont {Castro},\ and\ \citenamefont {Hung}}]{Belin:2015jpa}%
  \BibitemOpen
  \bibfield  {author} {\bibinfo {author} {\bibfnamefont {A.}~\bibnamefont
  {Belin}}, \bibinfo {author} {\bibfnamefont {A.}~\bibnamefont {Castro}},\ and\
  \bibinfo {author} {\bibfnamefont {L.-Y.}\ \bibnamefont {Hung}},\ }\href
  {https://doi.org/10.1007/JHEP11(2015)145} {\bibfield  {journal} {\bibinfo
  {journal} {JHEP}\ }\textbf {\bibinfo {volume} {11}},\ \bibinfo {pages}
  {145}},\ \Eprint {https://arxiv.org/abs/1508.01201} {arXiv:1508.01201
  [hep-th]} \BibitemShut {NoStop}%
\bibitem [{\citenamefont {Caputa}\ \emph
  {et~al.}(2017{\natexlab{a}})\citenamefont {Caputa}, \citenamefont {Kundu},
  \citenamefont {Miyaji}, \citenamefont {Takayanagi},\ and\ \citenamefont
  {Watanabe}}]{Caputa:2017urj}%
  \BibitemOpen
  \bibfield  {author} {\bibinfo {author} {\bibfnamefont {P.}~\bibnamefont
  {Caputa}}, \bibinfo {author} {\bibfnamefont {N.}~\bibnamefont {Kundu}},
  \bibinfo {author} {\bibfnamefont {M.}~\bibnamefont {Miyaji}}, \bibinfo
  {author} {\bibfnamefont {T.}~\bibnamefont {Takayanagi}},\ and\ \bibinfo
  {author} {\bibfnamefont {K.}~\bibnamefont {Watanabe}},\ }\href
  {https://doi.org/10.1103/PhysRevLett.119.071602} {\bibfield  {journal}
  {\bibinfo  {journal} {Phys. Rev. Lett.}\ }\textbf {\bibinfo {volume} {119}},\
  \bibinfo {pages} {071602} (\bibinfo {year} {2017}{\natexlab{a}})},\ \Eprint
  {https://arxiv.org/abs/1703.00456} {arXiv:1703.00456 [hep-th]} \BibitemShut
  {NoStop}%
\bibitem [{\citenamefont {Caputa}\ \emph
  {et~al.}(2017{\natexlab{b}})\citenamefont {Caputa}, \citenamefont {Kundu},
  \citenamefont {Miyaji}, \citenamefont {Takayanagi},\ and\ \citenamefont
  {Watanabe}}]{Caputa:2017yrh}%
  \BibitemOpen
  \bibfield  {author} {\bibinfo {author} {\bibfnamefont {P.}~\bibnamefont
  {Caputa}}, \bibinfo {author} {\bibfnamefont {N.}~\bibnamefont {Kundu}},
  \bibinfo {author} {\bibfnamefont {M.}~\bibnamefont {Miyaji}}, \bibinfo
  {author} {\bibfnamefont {T.}~\bibnamefont {Takayanagi}},\ and\ \bibinfo
  {author} {\bibfnamefont {K.}~\bibnamefont {Watanabe}},\ }\href
  {https://doi.org/10.1007/JHEP11(2017)097} {\bibfield  {journal} {\bibinfo
  {journal} {JHEP}\ }\textbf {\bibinfo {volume} {11}},\ \bibinfo {pages}
  {097}},\ \Eprint {https://arxiv.org/abs/1706.07056} {arXiv:1706.07056
  [hep-th]} \BibitemShut {NoStop}%
\bibitem [{\citenamefont {Caputa}\ \emph {et~al.}(2021)\citenamefont {Caputa},
  \citenamefont {Das},\ and\ \citenamefont {Das}}]{Caputa:2021pad}%
  \BibitemOpen
  \bibfield  {author} {\bibinfo {author} {\bibfnamefont {P.}~\bibnamefont
  {Caputa}}, \bibinfo {author} {\bibfnamefont {D.}~\bibnamefont {Das}},\ and\
  \bibinfo {author} {\bibfnamefont {S.~R.}\ \bibnamefont {Das}},\ }\href@noop
  {} {\  (\bibinfo {year} {2021})},\ \Eprint {https://arxiv.org/abs/2111.04405}
  {arXiv:2111.04405 [hep-th]} \BibitemShut {NoStop}%
\bibitem [{\citenamefont {Banerjee}\ and\ \citenamefont
  {Paul}(2017)}]{Banerjee:2015coc}%
  \BibitemOpen
  \bibfield  {author} {\bibinfo {author} {\bibfnamefont {S.}~\bibnamefont
  {Banerjee}}\ and\ \bibinfo {author} {\bibfnamefont {P.}~\bibnamefont
  {Paul}},\ }\href {https://doi.org/10.1007/JHEP02(2017)043} {\bibfield
  {journal} {\bibinfo  {journal} {JHEP}\ }\textbf {\bibinfo {volume} {02}},\
  \bibinfo {pages} {043}},\ \Eprint {https://arxiv.org/abs/1512.02232}
  {arXiv:1512.02232 [hep-th]} \BibitemShut {NoStop}%
\bibitem [{\citenamefont {Belin}\ \emph {et~al.}(2021)\citenamefont {Belin},
  \citenamefont {Myers}, \citenamefont {Ruan}, \citenamefont {S\'arosi},\ and\
  \citenamefont {Speranza}}]{Belin:2021bga}%
  \BibitemOpen
  \bibfield  {author} {\bibinfo {author} {\bibfnamefont {A.}~\bibnamefont
  {Belin}}, \bibinfo {author} {\bibfnamefont {R.~C.}\ \bibnamefont {Myers}},
  \bibinfo {author} {\bibfnamefont {S.-M.}\ \bibnamefont {Ruan}}, \bibinfo
  {author} {\bibfnamefont {G.}~\bibnamefont {S\'arosi}},\ and\ \bibinfo
  {author} {\bibfnamefont {A.~J.}\ \bibnamefont {Speranza}},\ }\href@noop {} {\
   (\bibinfo {year} {2021})},\ \Eprint {https://arxiv.org/abs/2111.02429}
  {arXiv:2111.02429 [hep-th]} \BibitemShut {NoStop}%
\bibitem [{\citenamefont {Liu}\ \emph {et~al.}(2021)\citenamefont {Liu},
  \citenamefont {Lyu},\ and\ \citenamefont {Raju}}]{Liu:2021hap}%
  \BibitemOpen
  \bibfield  {author} {\bibinfo {author} {\bibfnamefont {Y.}~\bibnamefont
  {Liu}}, \bibinfo {author} {\bibfnamefont {H.-D.}\ \bibnamefont {Lyu}},\ and\
  \bibinfo {author} {\bibfnamefont {A.}~\bibnamefont {Raju}},\ }\href@noop {}
  {\  (\bibinfo {year} {2021})},\ \Eprint {https://arxiv.org/abs/2108.04554}
  {arXiv:2108.04554 [hep-th]} \BibitemShut {NoStop}%
\bibitem [{\citenamefont {Leutheusser}\ and\ \citenamefont
  {Liu}(2021)}]{Leutheusser:2021qhd}%
  \BibitemOpen
  \bibfield  {author} {\bibinfo {author} {\bibfnamefont {S.}~\bibnamefont
  {Leutheusser}}\ and\ \bibinfo {author} {\bibfnamefont {H.}~\bibnamefont
  {Liu}},\ }\href@noop {} {\  (\bibinfo {year} {2021})},\ \Eprint
  {https://arxiv.org/abs/2110.05497} {arXiv:2110.05497 [hep-th]} \BibitemShut
  {NoStop}%
\end{thebibliography}%
\end{document}